\documentclass[]{interact}

\usepackage{epstopdf}
\usepackage[caption=false]{subfig}

\usepackage[numbers,sort&compress]{natbib}
\bibpunct[, ]{[}{]}{,}{n}{,}{,}
\makeatletter
\def\NAT@def@citea{\def\@citea{\NAT@separator}}
\makeatother

\usepackage{color} 
\usepackage{blindtext} 
\usepackage{bm} 
\usepackage{float}
\usepackage{hyperref}
\usepackage{textcomp}

\newcommand{\Secref}[1]{Sec.~\ref{#1}}
\newcommand{\Eq}[1]{Equation~\eqref{#1}}
\newcommand{\Fig}[1]{Figure~\ref{#1}}

\begin{document}

\title{Bose-Einstein condensates in microgravity and fundamental tests of gravity}

\author{
\name{Christian Ufrecht\textsuperscript{a}\thanks{CONTACT C. Ufrecht. Email: christian.ufrecht@gmx.de}, Albert Roura\textsuperscript{b} and Wolfgang P. Schleich\textsuperscript{a,b,c,d}}
\affil{\textsuperscript{a}Institut f{\"u}r Quantenphysik and Center for Integrated Quantum Science and Technology (IQ\textsuperscript{ST}), Universit{\"a}t Ulm, Albert-Einstein-Allee 11, D-89069 Ulm, Germany; \textsuperscript{b}Institute of Quantum Technologies, German Aerospace Center (DLR), S\"{o}flinger Stra\ss e 100, D-89077 Ulm, Germany; \textsuperscript{c}Hagler Institute for Advanced Study and Department of Physics and Astronomy, Institute for Quantum Science and Engineering (IQSE), Texas A{\&}M University, College Station, Texas 77843-4242, USA; \textsuperscript{d}Texas A{\&}M AgriLife Research, Texas A{\&}M University, College Station, Texas 77843-4242, USA}
}

\maketitle

\begin{abstract}
Light-pulse atom interferometers are highly sensitive to inertial and gravitational effects. As such they are promising candidates for tests of gravitational physics. In this article the state-of-the-art and proposals for fundamental tests of gravity are reviewed.
They include the measurement of the gravitational constant $G$, tests of the weak equivalence principle, direct searches of dark energy and gravitational-wave detection.
Particular emphasis is put on long-time interferometry in microgravity environments accompanied by an enormous increase of sensitivity. In addition, advantages as well as disadvantages of Bose-Einstein condensates as atom sources are discussed.
\end{abstract}


\newpage
\section{Introduction}
The wave nature of matter \cite{deBroglie} enables interferometry with massive particles such as electrons, neutrons, atoms, or molecules. In a matter-wave interferometer the wave function of a particle is split into two components that follow different paths, are subsequently redirected and finally recombined. As in conventional interferometry with light, the difference between the phases accumulated along the two branches of the interferometer can be obtained from the interference of the two components, and the quantity responsible for this phase shift can be inferred.

In contrast to interferometry with light, matter-wave interferometers are far more sensitive to gravitational effects due to the much longer interrogation times that are possible, which makes matter-wave interferometers very promising instruments for fundamental tests of gravity.

This review presents a selection of proposals and experiments for testing gravitational physics with light-pulse atom interferometers that can not only surpass classical tests in accuracy, but also constitute promising ways for probing regions of parameter space inaccessible to classical experiments.
Over the years a number of excellent reviews  \cite{ReviewBongs_2019, Review_Barrett2016, Kleinert,Review_Geiger2020,InterfaceGravity, Hogan, Review_Tino2020} with different emphasis and scope have been published. The present article focuses on a non-technical introduction to the theory and experimental techniques for fundamental tests of gravity with light-pulse atom interferometry. 

The article is organized as follows.
In  \Secref{Beam splitters and mirrors made out of light} the basic diffraction mechanisms  used to realize beam splitters and mirrors are introduced. From the interferometer phase shift, derived in \Secref{Atom interferometry for precision gravitational measurements}, the high sensitivity to inertial effects and gravity is subsequently inferred, and the main challenges as well as possible solutions are investigated. In \Secref{Long-time atom interferometry and associated challenges} various approaches to long-time atom interferometry are presented. The resulting high sensitivity is crucial for the fundamental tests of gravity outlined in \Secref{Fundamental gravitational measurements}.
In this respect, the extraordinary coherence properties and low expansion rate of Bose-Einstein condensates (BECs) make them an ideal atom source for long-time interferometry. The advantages and disadvantages of their use for atom interferometry are discussed in \Secref{BEC}.
Moreover, key mitigation techniques for major systematic effects are presented in \Secref{Mitigation techniques}.
Finally, we conclude with a brief outlook in \Secref{Outlook}.

\section{Beam splitters and mirrors made out of light}
\label{Beam splitters and mirrors made out of light}
Historically, evidence for wave-like behavior of matter was first found in electron experiments \cite{DavisonGermer} but soon also for molecules \cite{Molecules} and neutrons \cite{neutrons}. Interferometry with the latter allowed
 to measure the phase shift induced by the gravitational field of Earth \cite{ColellaOverhauser,SyntheticCrystal} for the first time with matter-wave interferometers. The next step was to perform interferometry with neutral atoms, which in contrast to neutrons are much easier to produce and in contrast to electrons are less sensitive to electromagnetic stray fields. Initially, experiments were performed with double slits \cite{DoubleSlit} and material gratings \cite{Keith1991}, but  it was soon realized that the internal structure of atoms provides excellent means for precise control with laser light. 
Profiting from these pioneering experiments and from advance in quantum optics, the beam splitters and mirrors were replaced by light gratings made out of standing waves \cite{Bragg3,Giltner1995,Rasel1995} or laser beams driving optical transitions \cite{Internal_state_Labeling,Riehle1991}.

\begin{figure}[ht]
	\begin{center}
		\includegraphics[width=\linewidth]{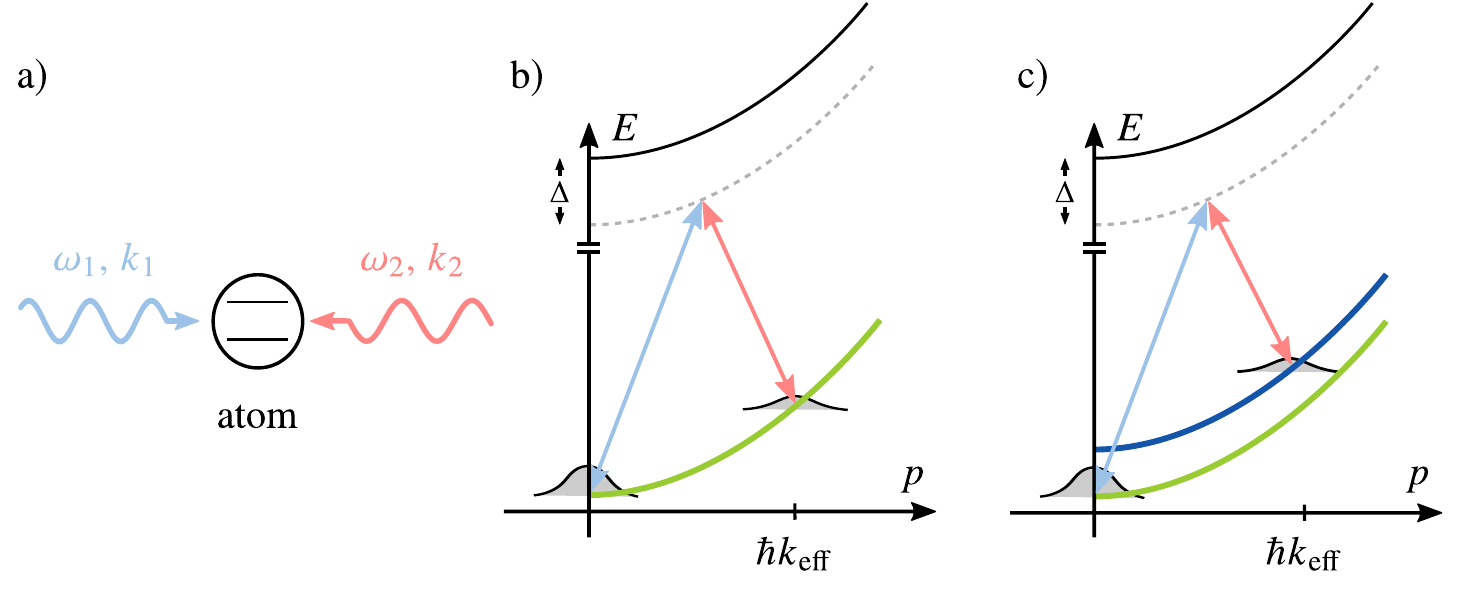}
		\caption{a) The diffraction process is realized by two counterpropagating laser beams of frequency $\omega_1$ and $\omega_2$ and the respective wave numbers $k_1$ and $k_2$. b) In Bragg diffraction a photon from the blue laser is absorbed and subsequently emitted by stimulated emission in direction of the red laser. The atom therefore gains the momentum $\hbar k_\mathrm{eff}=\hbar (k_1+k_2)$. The process is driven via a far-detuned virtual auxiliary state (black curve) back to the initial state of the atom. c) In Raman diffraction the frequencies of the lasers are chosen so that the atom is excited into a higher-energy state. Therefore, in contrast to Bragg diffraction in Raman diffraction the momentum $\hbar k_\mathrm{eff}$ is transferred together with a change of the internal state. 
		}
		\label{fig:RamanBraggMZ}
	\end{center}
\end{figure}

Alternatively, employing short laser pulses \cite{KasevichChu} provided more flexibility and pioneered the field of light-pulse atom interferometry, which will be at the focus of this review. As shown in  \Fig{fig:RamanBraggMZ} a), in a light-pulse atom interferometer beam splitters and mirrors are typically realized by subjecting the atoms to counterpropagating laser beams, driving a two-photon transition via a far-detuned auxiliary state. 
Intuitively, the process can be understood as the absorption of a photon associated with a recoil in direction of the first laser beam and a subsequent recoil in the same direction by stimulated emission  in direction of the second laser. Consequently, the atom gains the momentum $\hbar k_\mathrm{eff}$ with $k_\mathrm{eff}=k_1+k_2$ and the energy $\hbar \Delta \omega$ with $\Delta \omega=\omega_1-\omega_2$
where $k_1$, $k_2$ and $\omega_1$, $\omega_2$ are the wave numbers and frequencies of the laser pair. Therefore, during a pulse at time $t$ that transfers momentum $\hbar k_\mathrm{eff}$ to the atomic wave packet, the following laser phase is imprinted:
\begin{equation}
\label{Laserphase}
    \phi_\mathrm{L}(t)=k_\mathrm{eff} z(t)-\Delta \omega \, t -\Delta\varphi ,
\end{equation}
where $z(t)$ is the mean position of the wave packet
and the relative phase $\Delta\varphi = \varphi_1 - \varphi_2$ involves the difference between the phase terms of the two counter-propagating laser beams that are spatially and time-independent.

There are two different diffraction mechanisms commonly employed: Bragg diffraction \cite{Bragg1,Bragg2,Bragg3,Bragg4} and Raman diffraction \cite{KasevichChu, Internal_state_Labeling}. See Ref.~\cite{Hartmann1} for a comparison in terms of efficiency. In Bragg diffraction (\Fig{fig:RamanBraggMZ} b)) the atom returns via stimulated emission back to the original internal state while in Raman diffraction (\Fig{fig:RamanBraggMZ} c)) a different internal state is selected (typically another hyperfine state).
The frequency difference required between the laser pair depends on which of these two processes is driven and is determined by energy and momentum conservation.
By choosing the interrogation time and laser intensity appropriately, beam splitters and mirrors can be realized. A $\pi/2$ pulse, or beam splitter, sets the initial wave packet with mean momentum $p_0$ into an equal superposition of two components, one with mean momentum $p_0$ and another one with $p_0+\hbar k_\mathrm{eff}$.  In contrast, a $\pi$ pulse, or mirror, inverses the momenta of the two components. In case of Raman diffraction each momentum transfer is additionally associated with a change of the internal state.

\begin{figure}[ht]
	\begin{center}
		\includegraphics[width=\linewidth]{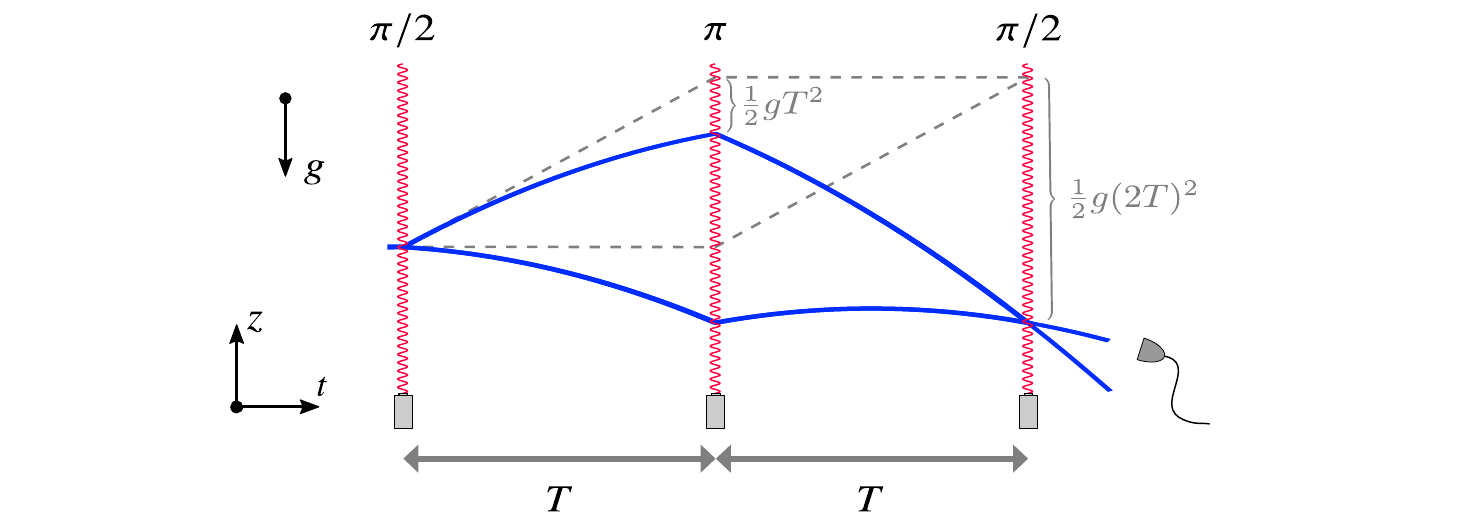}
		\caption{Space-time diagram of a Mach-Zehnder (MZ) interferometer. An MZ interferometer consists of two $\pi/2$-pulse beam splitters and a $\pi$ pulse (mirror) separated by the free interrrogation time $T$. Initially, the wave function of the atoms is split into two components with momentum transfer $\hbar k_\mathrm{eff}$ on one of them. Due to the momentum difference, the two components subsequently spatially separate, constituting the two branches of the interferometer. At the final $\pi/2$ pulse the two components are interfered and the phase shift between the two branches can be inferred by measuring the number of atoms leaving each exit port. In a gravitational field with acceleration $g$ the atoms are in free fall between the laser pulses, the trajectories are shown by the blue solid parabolas and are compared to the grey dashed lines corresponding to zero gravity. As shown in \Secref{Inertial sensors based on atom interferometry}, the phase shift due to the gravitational field in a MZ gravimeter is $\delta\phi=-k_\mathrm{eff}gT^2$.
		}
		\label{fig:MZInterfermeter}
	\end{center}
\end{figure}

A light-pulse interferometer consists of an alternating sequence of beam splitters, mirrors and free evolution in between.
One common type of interferometer scheme used for many applications due to its simplicity is the Mach-Zehnder (MZ) interferometer displayed in \Fig{fig:MZInterfermeter}.
In an MZ interferometer a $\pi/2$ pulse coherently splits the initial wave function into two components which subsequently spatially separate due to their momentum difference $\hbar k_\mathrm{eff}$, constituting the two branches of the interferometer. After time $T$ the two components are redirected by a $\pi$ pulse and finally interfered by a second $\pi/2$ pulse after another time $T$ has passed.
Subsequently, the phase shift between the two branches is read out from the fraction of atoms detected at each exit port of the interferometer through fluorescence or absorption imaging.

To relate the measurement result to a theoretical model, the interferometer phase shift can be calculated within several levels of approximation.  
In the first step, the far-detuned auxiliary state can be adiabatically eliminated \cite{AdiabaticElimi,AdiabaticElimi2, AdiabaticElimi3}, leading to a reduced description in which the atom couples to an effective light field with wave number $k_\mathrm{eff}$ and frequency $\Delta \omega$.
Next, the center-of-mass motion of the atom and internal degrees of freedom corresponding to the two remaining internal states in case of Raman diffraction are separated. In contrast, for Bragg diffraction well separated momentum classes are introduced \cite{AdiabaticElimiEnno} centered around multiples of $\hbar k_\mathrm{eff}$ which are coupled only during the lasers are on.
Assuming near-resonance diffraction and negligible occupation of higher momentum classes in case of Bragg diffraction (two-level approximation), further reduction to a path-dependent description is possible \cite{Marzlin}, where the probability for an atom to leave at the first exit port is given by
\begin{equation}
    P=\frac{1}{2}\big(1+\mathfrak{Re}\big\langle \hat{U}^{(l)\dagger} \hat{U}^{(u)}\big\rangle\big)=\frac{1}{2}(1+C \cos{\delta\phi})\,.
    \label{eq:AI_probability}
\end{equation}
In this expression the expectation value with respect to the initial wave function defines the phase shift $\delta\phi$ and contrast $C$ of the interferometer.  The center-of-mass time-evolution operators $\hat{U}^{(\alpha)}$ including the momentum transfer and laser-phase imprint generate the motion along the upper and lower branch ($\alpha=u, l$).
Similarly, the probability for the second exit port is given by an analogous expression but with the opposite sign for the term proportional to $\cos \delta\phi$. Therefore, by measuring the fraction of atoms detected at each exit port, one can infer the phase shift $\delta\phi$.

Many different methods have been developed to calculate phase shift and contrast at this stage, each of them with particular advantages and disadvantages regarding range of application and computational ease. Noteworthy are path-integral methods \cite{Tannoudji,Borde2}, calculations in phase space \cite{Dubetsky2006,InterfaceGravity}
and representation-free descriptions on the operator level \cite{RedshiftSchleich,Marzlin,Kleinert} in terms of comoving frames \cite{Hogan,OvercommingLoss,ZellerPhd,UfrechtPhd} or perturbation theory \cite{UfrechtPhd, PAI} as well as a relativistic description in curved spacetime \cite{UGRRoura,Dimopoulos2008a}.
In the following sections the description given so far is extended and the details needed to understand atom-interferometric fundamental test of gravity are elaborated.

\section{Atom interferometry for precision gravitational measurements} 
\label{Atom interferometry for precision gravitational measurements}
Since the first application of light-pulse atom interferometers as gravity sensor in 1991 \cite{DroppingAtoms} enormous effort has been devoted to increasing their sensitivity and characterizing systematic effects.
Based on these improvements, today's light-pulse atom interferometers are highly sensitive inertial sensors. Indeed, absolute gravimeters \cite{Altin_2013, PetersMetrologia, Merlet_2010, SYRTE, AbsGravity1, AbsGravity2,AtomChipGravi} used to measure the local gravitational acceleration have demonstrated sensitivities of $4.2  \times  10^{\text{-}8}\,\mathrm{m}\, \mathrm{s}^{\text{-} 2}\,\mathrm{Hz}^{\text{-}1/2}$   \cite{AbsGravity1}. In gyroscopy \cite{Gyroscopy1997, Gyroscopy2000, Gyroscopy2006, Gyroscopy2006_2, Gyroscopy2015, Gyroscopy2016, Gyroscopy2018,Sagnac_Review}
a sensitivity of $3 \times 10^{\text{-}8}\mathrm{rad}\, \mathrm{s}^{\text{-}1} \mathrm{Hz}^{\text{-}1/2}$ has been reported \cite{Gyroscopy2018}. These sensitivities are comparable to those of the best classical gravimeters \cite{ClassicalGravimetry} and gyroscopes \cite{ComparisonGyroscopy}.
The decisive benefit of atom interferometers, however, is their high stability due to low drift rates.
Indeed in gyroscopy a record stability  of $3 \times  10^{\text{-}10}\,\mathrm{rad}\, \mathrm{s}^{\text{-} 1}$ \cite{Gyroscopy2018} was reached.
For applications in navigation, geodesy \cite{GravimetryOnShip,AIOnPlane1}, seismometry and other geophysical studies, compact and mobile devices have been developed \cite{MobileDevice,CompactDevice, MobileDevice2018} designed to leave the laboratories. 
In the following, in \Secref{Inertial sensors based on atom interferometry} the phase shifts measured by a gravimeter and a gyroscope is motivated for an idealized experiment.
In \Secref{Challenges and possible solutions} challenges in more realistic situations and possible solutions are discussed.

\subsection{Inertial sensors based on atom interferometry} 
\label{Inertial sensors based on atom interferometry} 
 In this section the principles underlying the operation of light-pulse atom interferometers as  gravimeters and gyroscopes are illustrated.
These devices measure the acceleration and rotation rate of the frame defined by the laser beams.
They rely on the fact that neutral atoms in magnetically insensitive states are excellent inertial references and on the interferometer phase shift being directly related to the relative acceleration between the atoms and the laser wave fronts.
The latter point can be easily understood by considering a freely falling frame, where the atoms move with constant velocity between the laser kicks.
In particular, for a linear gravitational potential the phase difference between upper and lower branch can be attributed to the laser phase imprinted on the wave packets at each laser pulse. Consequently, by comparing to \Fig{fig:MZInterfermeter}, the phase shift for an MZ interferometer reads $\delta\phi=\phi_\mathrm{L}(0)-2\phi_\mathrm{L}(T)+\phi_\mathrm{L}(2T)$. In the freely falling frame the frequency $\Delta\omega$ appearing in \Eq{Laserphase} for the laser phase is Doppler shifted so that $\phi_\mathrm{L}(t)=k_\mathrm{eff} z(t)-\Delta \omega\, t + \bm{k}_\mathrm{eff} \cdot \bm{a}\, t^2/2 - \Delta\varphi$ where $\bm{a}$ is the acceleration of the atoms with respect to the reference frame of the laser. For a MZ interferometer the spatial part of the laser phases cancels out
and so does $\Delta\varphi$ provided that the relative phase between the counter-propagating laser beams is properly stabilized during the whole interferometer sequence.
Thus, the phase shift becomes $\delta \phi = \bm{k}_\mathrm{eff} \cdot \bm{a}\,T^2$.
As the atoms accelerate relative to the laser wave fronts, they quickly leave resonance which can be compensated in an accelerometer  by dynamically modulating the  effective laser frequency with a chirp rate $\alpha$, so that $\Delta \omega\, t\rightarrow\Delta \omega\, t-\alpha t^2/2$. At resonance  $\alpha=-\bm{k}_\mathrm{eff} \cdot \bm{a}$ and the interferometer phase shift vanishes. However, due to Heisenberg's time-energy uncertainty relation and the finite pulse duration, this resonance has a finite width, so that beam splitters and mirrors are still possible close to the resonance condition.
The acceleration can then be read out from the phase shift
\begin{equation}
\label{MZPhase}
\delta\phi=(\bm{k}_\mathrm{eff} \cdot \bm{a} +\alpha) T^2
\end{equation}
by scanning over $\alpha$ at constant $T$ in repeated measurements. In a linear gravitational field an absolute gravimeter measures the local gravitational acceleration $\bm{a}=\bm{g}$.
Note that \Eq{MZPhase} holds in the limit of vanishing pulse duration and small corrections areise when taking into account their finite duration \cite{Antoine2003b,Cheinet2008}.
 
The measurement of rotations can be understood as follows. If the experimental set-up, including the laser beams, rotates with an angular velocity $\bm{\Omega}$, the atoms will experience in that frame a Coriolis acceleration $\bm{a} = 2 (\bm{v}_0 \times \bm{\Omega})$, where $\bm{v}_0$ is the initial velocity of the atoms. This acceleration gives rise to the following Sagnac phase shift:
\begin{equation}
\label{eq:rotation_phase}
\delta\phi=2 \bm{k}_\mathrm{eff} \cdot\left(  \bm{v}_0 \times \bm{\Omega} \right)T^2 ,
\end{equation}
which is obtained by substituting the Coriolis acceleration into \Eq{MZPhase}.
When both linear acceleration and rotations are present, there are further phase-shift contributions coupling $\bm{a}$ and $\bm{\Omega}$ \cite{Kleinert}.

  \subsection{Challenges and possible solutions}
 \label{Challenges and possible solutions}
 In the previous section the origin of the phase shift in a measurement of inertial effects was explained in an ideal experiment. Of course in reality the situation is less idealized and several noise sources limit the precision. In this subsection laser phase noise, impact of optical elements and vibration noise is briefly discussed.
 
 As discussed in the previous section each laser pulse imprints the  phase $-\Delta \omega t$ on the wave packet. Consequently, random fluctuations of the laser phase directly translate into the interferometer phase. By phase locking the lasers,  their frequency difference  can be controlled by referencing to an external stable frequency source \cite{LaserPhaseNoise,PetersMetrologia,KasevichChu}.
 
Additional noise from collimators, beam shapers, optical fibers etc.~during preparation of the beams can be strongly suppressed  by feeding both beams into the vacuum chamber via the same optical components and retroreflecting them on a mirror \cite{PetersMetrologia}. The polarization of the two beams is rotated by $\pi/2$ by passing two times through a quarter-wave plate in front of the mirror. Choosing the appropriate initial polarization, only pairs of beams with different laser frequency interact and further undesired scattering processes are excluded.
In a gravimeter configuration only one pair of beams drives transitions as the other is strongly Doppler detuned due to the free fall of the atoms. 
In zero-gravity environment, in contrast, both laser pairs interact with the atoms, resulting in Double diffraction \cite{DoubleRaman, AdiabaticElimiEnno, DoubleRamanGravimeter,DoubleBragg}. 
 
The remaining noise from vibrations severely limits state-of-the-art absolute gravimetry. While vibrations of the optical components and the laser system are rejected when the laser beams travel together to the injection point, parasitic vibrations of the mirror constitute a major source of noise (which is not rejected since only one beam is reflected off the mirror).
The distance travelled by this beam varies between different pulses due to the vibrational motion of the mirror.
In the quasistationary approximation \cite{Vibrations}
this shifts the laser-phase contribution $\Delta\varphi_j$ for each pulse by $-k_\mathrm{eff} z_\mathrm{m} (t_j)$, where $z_\mathrm{m} (t_j)$ is the position of the mirror at the time $t_j$ of the $j$th laser pulse, and results in the following additional contribution to the total phase-shift:
\begin{equation}
\label{PhaseVibration}
    \delta\phi_{\text{vib}}=-k_\mathrm{eff}[z_\mathrm{m} (0)-2z_\mathrm{m} (T)+z_\mathrm{m} (2T)]\,.
\end{equation} 

The effect of vibrations can be mitigated with the help of vibration isolation platforms \cite{VibrationIsolation1,VibrationIsolation2} in combination with additional active feedback control  \cite{VibrationIsolation3}.
The residual effect of vibrations imposes the current limit on the sensitive for absolute measurements of gravity.  

Instead, when two gravity measurements are performed simultaneously with two interferometers and the laser beams are reflected off the same mirror, the phase shifts from vibrations cancel in the differential phase shift. This suppression in a differential measurement can be exploited for fundamental tests of gravity. Here, two experimental arrangements suggest themselves. (i) In gradiometry the local gravitational acceleration is measured at two different locations allowing the detection of spatial variations of gravity with applications for the measurement of Newton's gravitational constant $G$ as discussed in \Secref{Measurements of Newton's gravitational constant G}.
(ii) In tests of the universality of free fall, discussed in \Secref{Tests of the weak equivalence principle}, the two interferometers are superimposed and the differential acceleration of two different atomic species is measured.

Before the experimental setups are explained in more detail, possible strategies to increase the sensitivity of atom interferometers are illustrated in the next section.

\section{Long-time atom interferometry and associated challenges} 
\label{Long-time atom interferometry and associated challenges} 
In a measurement of the local gravitational acceleration $g$ with a gravimeter the noise in the value of $g$ follows from \Eq{MZPhase} by error propagation as
\begin{equation}
\label{SensitivityGravity}
    \Delta g=\frac{\Delta \phi}{k_\mathrm{eff}  T^2}
\end{equation}
 if  $\bm{k}_\mathrm{eff}$ orthogonal to $\bm{a}=\bm{g}$ and assuming the values of $k_\mathrm{eff}$ and $T$ to be known. The phase-shift uncertainty $\Delta \phi$  consists of two sources: (i) Uncontrollable phase shifts, for instance from laser phase noise, vibration noise etc. and (ii) the uncertainty by which the phase shift itself can be extracted from the interferometer signal.  For uncorrelated particles the latter is limited by shot noise which is given by $1/\sqrt{N}$ where $N$ is the number of particles per shot. This noise term originates from the fluctuation of $\sqrt{N}$ particles over the two exit ports in case of independent particles. For subsequent measurements uncorrelated random fluctuations with zero bias additionally decrease with $1/\sqrt{n}$ where $n$ is the number of shots.

The inverse scaling with $k_\mathrm{eff}T^2$ in \Eq{SensitivityGravity} suggests two  strategies to decrease $\Delta g$: (i) Large-momentum transfer techniques (LMT) resulting in large effective wave vectors $\bm{k}_\mathrm{eff}$ and (ii) increasing the interrogation time $T$. 
With the former strategy considerable progress has been achieved over recent years with higher-order Bragg and Raman diffraction \cite{highorderBragg, Hartmann2}, sequential Bragg diffraction   \cite{SequentialBragg,SequentialBraggRecord}, Bloch oscillations \cite{BlochOscillation,Bloch1996,DeltaKick2013} and combinations  \cite{Bloch-Bragg} which has resulted in an enormous increase of the space-time area enclosed by the branches of the interferometer \cite{LMT1, LMT2, LMT3,SequentialBragg}.
However, spurious diffraction phases \cite{DiffractionPhases2003} and parasitic interferometers \cite{SpuriousInterferometers2016} arise which are hard to control.

The second strategy is based on increasing the interrogation time $T$ which might even be more effective due to the quadratic scaling of $T$ in \Eq{SensitivityGravity}. On Earth  the free-fall time of the atoms is limited by the size of the apparatus. Still,  total interferometer times of up to 2.3s have been realized in the Stanford 10-meter tower \cite{FirstPublicationStanfordTower}  by launching the atoms initially.

This limitation can be overcome by relocating the setup to microgravity environments which will be briefly discussed in the following.

\subsection*{Microgravity platforms} 
Experiments in several different microgravity facilities have been performed.
In parabola flights 20s of zero $g$ were realized \cite{AIOnPlane1, AIOnPlane2} 
and a first proof-of-principle test of the universality of free fall in microgravity was performed
in which the influence of the strong vibrational background on a plane could be rejected to some extent.
Much quieter environments are provided in droptower facilities as for example in Bremen \cite{DropTower} where experiment times of about 5s in the drop mode and almost 10s in the catapult mode are available \cite{VanZoest2010,Muntiga2013}.
However, repetition rates are low in the droptower.
Much higher repetition rates and free-fall times of about $4$s can be reached in the newly commissioned Einstein Elevator in Hanover \cite{EinsteinElevator}.
The proof-of-principle experiment in a sounding rocket \cite{SpaceBorneBEC, MaiusInterferometry}
 offers valuable insight into challenges associated with future experiments in space such as on the International Space Station (ISS) \cite{CAL2018, BECAL,WEPProposalISS} or even in dedicated satellite missions \cite{WEPSatelliteProposal,STEQUEST} with in principle unlimited interferometer time.

The combination of differential measurements discussed in \Secref{Challenges and possible solutions} which large common-mode suppression and experiments in microgravity with large interferometer times allows to reach high sensitivities  for experimental tests of fundamental properties of gravity which will be discussed in the following section.

\begin{figure}[h]
	\begin{center}
		\includegraphics[width=\linewidth]{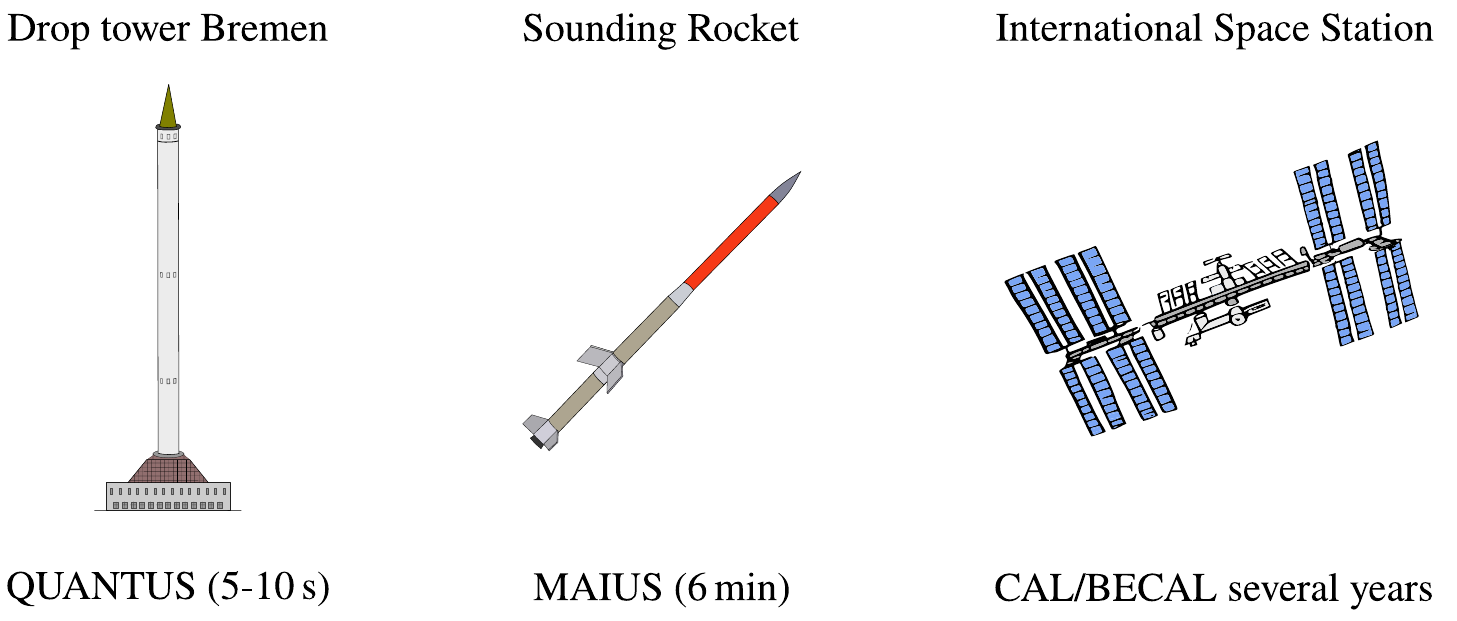}
		\caption{
		The drop tower at the Center for Applied Space Technology and Microgravity (ZARM) in Bremen provides a platform for high-quality weightlessness of almost 10 seconds inside an evacuated tube.
		In the 2017 sounding-rocket mission MAIUS-I 6 min of in-space flight have been realized and the production and manipulation of  BECs as well as basic interferometric principles have been demonstrated. The International Space Station in principle provides unlimited interferometer times in microgravity. In NASA's Cold Atom Laboratory (CAL) first interferometer experiments have been performed. Based on the experiences from CAL, the Bose-Einstein Condensate and Cold Atom Laboratory (BECCAL) will be installed in the coming years and operated by a collaboration between NASA and DLR.}
		
		\label{fig:Microgravity}
	\end{center}
\end{figure}

\section{Fundamental gravitational measurements}
\label{Fundamental gravitational measurements}
In the previous sections the basic ideas of light-pulse atom interferometry were introduced and challenges together with possible solutions were illustrated. Building on these principles, in this section measurements and proposals for probing fundamental gravitational physics are illustrated with focus placed on experiments particularly suited for operation in microgravity and in a differential mode. This includes measurements of Newton's constant $G$, tests of the weak equivalence principle, dark energy and gravitational-wave detection. 

Other tests of fundamental physics not discussed in this review are measurements of the fine structure constant with unprecedented accuracy \cite{FineStructure1,FineStructure2,FineStructure3,Morel2020} as well as measurements of special-relativistic time dilation \cite{SMI, TwinParadox} and the gravitational redshift \cite{UGRRoura, UGRUfrecht,Roura2020b, DiPumpo2021} using quantum-clock interferometry \cite{Sinha2011, Zych2011}.

\subsection{Measurements of Newton's gravitational constant}
\label{Measurements of Newton's gravitational constant G}
Newton's constant $G$ is one of the least accurately determined constants of nature but even at the present level of accuracy conflicting values have been reported \cite{ComparisonG}. While classical measurements have been performed mainly with torsion balances or pendulums \cite{ComparisonG}
and their working principle is very similar to the first measurement of $G$ performed by Cavendish more than two centuries ago,
an atom-interferometric measurement provides a complementary experiment of completely different type which might help reveal unaccounted  systematic effects and resolve the dissatisfying inconsistencies.
The important role $G$ takes in astrophysical and geophysical calculations makes a more accurate value of the constant an important goal.

The gravitational constant $G$ can be determined by measuring the gravitational field generated by a well-characterized source mass, which can in turn be obtained from the acceleration experienced by a test mass close to the source mass. In particular, one can make use of atom interferometry to measure the acceleration of the atoms caused by a nearby macroscopic mass.
However, a precise measurement is very challenging because the gravitational force is rather weak and experiments cannot be shielded from the gravitational environment. This means that one needs to separate the small signal of interest from the much larger acceleration associated with Earth's gravitational field, which can be achieved by performing a differential measurement of two spatially separated atom interferometers. In that way, the uniform part of Earth's gravity field cancels out. Besides the inhomogeneous gravitational field generated by the source mass, Earth's gravity gradient will also contribute to the differential measurement, but the two contributions can be separated by repeating the measurement for different positions of the source mass.

\begin{figure}[h]
	\begin{center}
		\includegraphics[width=\linewidth]{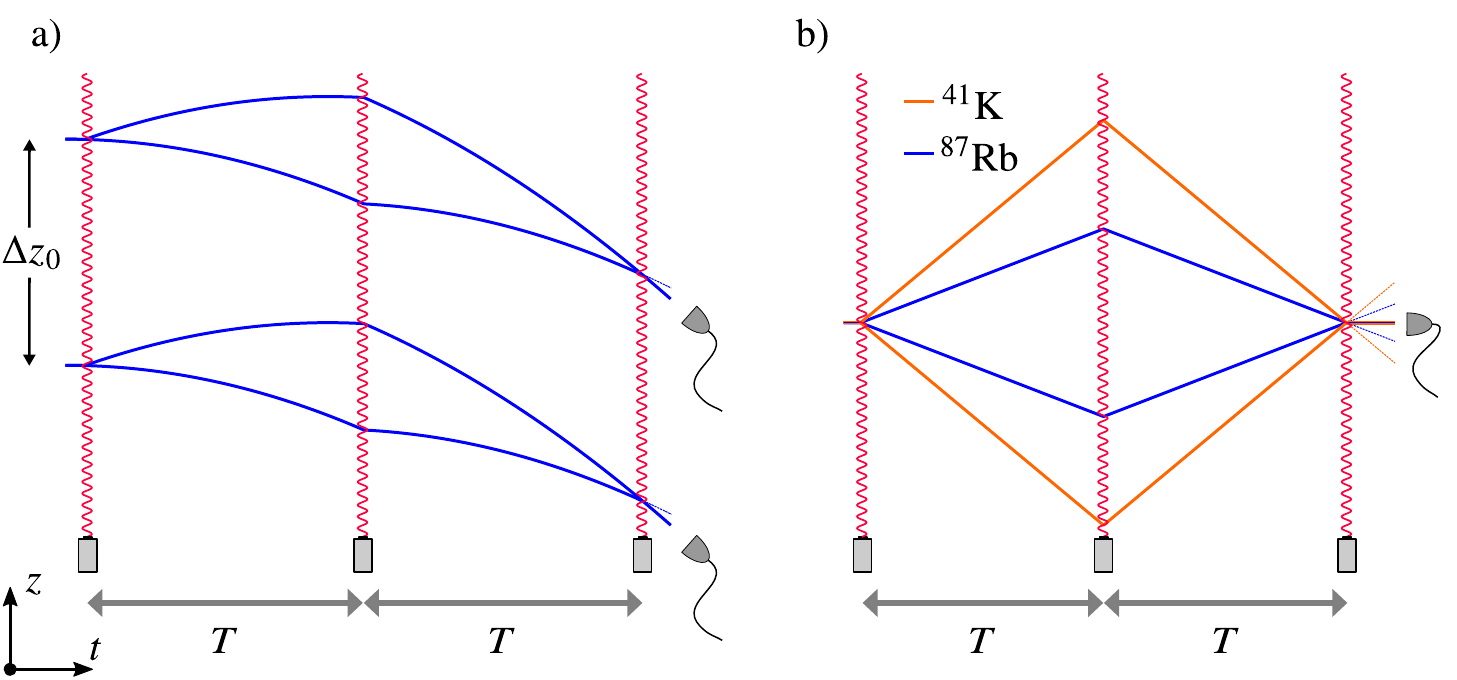}
		\caption{a) Gradiometry configuration. Two gravimeters (see \Secref{Atom interferometry for precision gravitational measurements}) separated by a distance $\Delta z_0$ are  simultaneously interrogated with the same laser pair. In contrast to absolute gravimetry, where the absolute value of the gravitational field is detected, in this gradiometry setup spatial variations over the distance $\Delta z_0$  can be extracted from the differential phase shift. Note that the branch separation is not to scale in the figure. Indeed,  as an example, for an interferometer with $^{87}$Rb atoms, $T=1$s and a momentum transfer $\hbar k_\mathrm{eff}$ corresponding to twice the single-photon recoil, the branch separation would be of the order of one centimeter while $\Delta z_0$ may typically be of the order of a meter.
		b) Test of the weak equivalence principle. Two interferometers with different atomic species e.g.~$^{87}$Rb and $^{41}$K are superimposed.
		The relative acceleration between the atoms can be obtained from the differential phase shift 
		with nonzero result in case the WEP is violated. Note the separation between the potassium and rubidium trajectories due to their different mass and different laser frequencies required to drive the respective internal transitions.
		The trajectories in the figure are to scale for the parameters in the space-borne proposal of Ref.~\cite{WEPSatelliteProposal} with $T=20$s. The parameters lead to a maximal branch separation of about $1$m for $^{41}$K and $0.5$m for the $^{87}$Rb atoms.}
		\label{fig:Gradiometry}
	\end{center}
\end{figure}

More specifically, a differential measurement of this kind can be achieved in a gradiometry configuration where two interferometers are placed at different locations \cite{Gradiometrie1, Gradiometrie5, Gradiometrie6, Gradiometrie4, Gradiometry7} as shown in \Fig{fig:Gradiometry}a.
Both interferometers are interrogated with the same laser beam retroreflected off a common mirror. As a consequence, vibration and laser phase noise, which would otherwise mask the signal of interest, are common to both interferometers and largely cancel out in the differential measurement.
The gravitational potential on Earth's surface can be decomposed into the dominant linear part $mg z$ with nonlinear perturbations on top. Since the acceleration exerted on a test particle by the former is constant over the whole gradiometry setup, 
 the differential phase shift between the interferometers only depends on the latter, the spatial variations of the gravitational field.
This phase-shift difference can be extracted by plotting the signal of the interferometers against each other. As a function of the chirp rate of the Raman lasers the signals parametrize an ellipse \cite{EllipseFitting} from which the differential phase shift can be extracted by a least-square fit even in noisy environments.
In fact, due to vibration noise, the phase-shift contribution $\delta\phi_\mathrm{vib}$ given by \Eq{PhaseVibration} will typically take values randomly distributed in the whole range $(-\pi,\pi)$ for this kind of experiments, giving rise to data points randomly distributed along the ellipse even for a fixed chirp rate.

The practicability of the experimental setup explained above  for a measurement of Newton's constant was shown in proof-of-principle experiments \cite{Newton1,Newton2,Newton4}. In the first high-precision measurement of $G$ using atom interferometry, a relative uncertainty of 150ppm  was reported \cite{Newton3} deviating 1.5 combined standard deviations from the CODATA recommended value \cite{CODATANewton} at that time but also was the first value measured by a new type of experiment.
In Ref.~\cite{Newton3} the atoms of both interfermeters are launched simultaneously close to two annular test masses at variable position made out of 516kg of tungsten. The limiting systematics of the experiment were identified as mounting errors and inhomogeneities in the density of the test mass as well as uncertainties in the initial position of the atoms and the large cloud size with further improvement expected in the future \cite{FuturePerspective}. Dependence of the phase shift on initial kinematics can be strongly suppressed by the mitigation techniques explained in \Secref{Rotations}  and \Secref{Gravity gradients}. The spread of the wave function can be drastically reduced in both position and momentum space by employing BECs as atom sources as explained in \Secref{BEC}.

In a similar setup Yukawa-type fifth forces were constrained \cite{Gradiometrie3}
with great potential improvement by local gravity measurements using Bloch oscillations in optical lattices close to a source mass \cite{BigG_Bloch, BigG_Bloch2,Fermi_TestingGravity}.

\subsection{Tests of the weak equivalence principle}
\label{Tests of the weak equivalence principle}

The weak equivalence principle, also referred to as universality of free fall (UFF), is an essential ingredient for the construction of general relativity (GR). It states that the acceleration experienced by a test particle in a gravitational field is independent of the particles's composition as long as it is kept together only by non-gravitational forces \cite{NonequivalenceOfEquivalence}. Being a necessary condition for GR, any violation of the weak equivalence principle would prove that GR is not the ultimate gravitational theory but needs to be modified in some respect. There are various possible origins of violations for example from modified gravity-matter coupling or standard-model extensions \cite{Damour_2012,MuellerEquivalence}.

Comparing the accelerations $g_1$ and $g_2$ experienced by two test particles, a natural measure of the size of possible violations is given by the Eötvös parameter
\begin{equation}
\label{Eotvosh}
    \eta_{1,2}=2\frac{g_1-g_2}{g_1+g_2}
\end{equation}
which vanishes if no violation is present. UFF tests aim at confining the value of $\eta_{1,2}$ and by that allow to set bounds on parameters of alternative gravitational theories. Vice versa once the rough form of couplings in a modified theory is evaluated, the best choice of test masses that maximize the Eötvös parameter can be predicted \cite{ALTSCHUL2015}. 

By comparing the acceleration of macroscopic test masses with torsion balances on Earth, the violation parameter was confined to the $10^{-13}$ level \cite{TorsionPendulum1,TorsionPendulum2,TorsionPendulum3} and recently $\eta=2\times10^{-14}$ consistent with no violation of the UFF was reported by the MICROSCOPE mission  \cite{MICROSCOPE_2017, Microscope_2019} in a satellite-based measurement in an orbit around Earth.

The advent of atom interferometry has enabled quantum tests of the weak equivalence principle
that contribute to broadening the range of possible materials employed as test masses.
Furthermore, in contrast to macroscopic objects, all atoms of the same isotope and in the same internal state are identical and are therefore less prone to certain kinds of systematic errors associated with macroscopic test masses in classical tests.

Using effective-absorption gratings of light, the first atom interferometric test was performed in 2004, which comparied the acceleration of $^{87}$Rb and $^{85}$Rb isotopes \cite{WEPStandingWave}. A few years later approximately the same accuracy was reached in the first light-pulse atom interferometer based test \cite {WEPBonnin}, later pushed further by another order of magnitude \cite{WEPIsotopes}. In addition, tests for probing gravity-spin coupling have been performed \cite{WEPSpinOrientations}.
The first test with different atomic species validated the weak equivalence principle at the  $10^{-7}$ level by comparing the accelerations of $^{87}$Rb and $^{39}$K atoms \cite{WEPSchlippert}. At about the same time, the acceleration of two Sr isotopes was compared by measuring the Bloch frequency in an optical lattice with the result $\eta<10^{-7}$ \cite{WEPGravitySpinCoupling}.
Recently, the UFF was verified at the $10^{-12}$ \cite{WEPKasevich} level for rubidium isotopes using light-pulse atom interferometers, therefore improving on earlier results by more than four orders of magnitude.

In a WEP test, as illustrated in \Fig{fig:Gradiometry} b), two interferometers involving different atomic species are operated simultaneously. However, in general not only the internal transitions used for the diffraction but also the mass of the two atomic species are different, leading to different recoil velocities.
The resulting separation between the trajectories of the two species can range from few millimeters in experiments on ground with different isotopes of the same chemical element up to about half a meter in proposals for satellite missions with different elements.

As explained in \Secref{Inertial sensors based on atom interferometry}, the phase shift accumulated by species $j$ is  $\delta\phi_j=k_{\mathrm{eff}}^{(j)}\,g_j T^2$+$\Delta \phi_j$. The first term originates from the linear gravitational potential where the species-dependent acceleration $g_j$ models possible violations of the UFF.
 Any additional phase-shift contribution is comprised in $\Delta \phi_j$. The differential effective acceleration
\begin{equation}
\label{WEPAcceleration}
    \bar{g}_1-\bar{g}_2=\frac{\delta\phi_1}{k_{\mathrm{eff}}^{(1)} T^2}-\frac{\delta\phi_2}{k_{\mathrm{eff} }^{(2)}T^2}=g_1-g_2+\left(\frac{\Delta\phi_1}{k_{\mathrm{eff}}^{(1)} T^2}-\frac{\Delta\phi_2}{k_{\mathrm{eff} }^{(2)}T^2}\right)
\end{equation}
is proportional to the Eötvös parameter defined in \Eq{Eotvosh} plus additional terms dependent on the spurious phase shifts $\Delta \phi_j$. The theoretical accuracy of the WEP test is then determined by the difference inside the brackets. The differential nature of the scheme leads to cancellation of many systematic effects like 
 leading-order phase shifts from Coriolis effects and gravity gradients. Also note that
 the position of the mirror is common to both species. Consequently, phase shifts from vibration noise which are proportional to $k_\mathrm{eff}^{(j)}$ as shown in \Eq{PhaseVibration}, largely cancel in \Eq{WEPAcceleration} .

Other systematic effects that act differently on both species and hence do not cancel in the differential phase shift need to be well controlled such as magnetic field gradients (second-order Zeeman effect \cite{MagneticShielding}) and temperature gradients in the setup (black-body radiation \cite{BBR1,BBR2}).
The most precise test performed so far with atom interferometry  \cite{WEPKasevich} is currently limited by two systematic effects. The first is due to kinematic differences of the initial atomic clouds. Mitigation strategies to reduce the impact of this effect on the phase shift are discussed in detail in \Secref{Mitigation techniques} and already are a major ingredient in the experimental setup of Ref.~\cite{WEPKasevich}. But even stronger mitigation will be required for tests beyond the current accuracy. The second uncertainty is due to the shift of the atomic levels during the lasers are turned on (AC-Stark shift). While in case of Bragg diffraction this effect ideally cancels separately for each species, laser intensity fluctuations over the separation of the branches induce a non-negligible phase shift associated with pulses that are resonant with only one of the two arms, which arise when employing multi-pulse sequences for larger momentum transfer.
Moreover, the inhomogeneous gravitational potentials induced by local mass sources surrounding the experiment lead to nonvanishing differential accelerations on the two species as they follow different trajectories.
This remaining bias in \Eq{WEPAcceleration} might limit future WEP tests on ground if the gravitational environment is not precisely characterized.

Ambitious proposals for further improving the current limits have been put forward for experiments on ground \cite{WEPProposalVLBAI, WEPProposalOnGround} and in space \cite{STEQUEST, WEPSatelliteProposal,Gauge}.
Experiments performed in satellites orbiting Earth \cite{WEPSatelliteProposal} can in particular profit from the quiet environment and a further mitigation technique, also discussed in \Secref{Mitigation techniques}.

\subsection{Dark energy}
\label{sec:dark_energy}

Cosmological observations have revealed that about 70\% of the energy density of the cosmos corresponds to \emph{dark energy}, a form of energy that cannot be accounted for by the Standard Model of particle physics and drives the current accelerated expansion of the universe \cite{Frieman2008,Weinberg2013}. Its defining feature is the large negative pressure, comparable in absolute value to its energy density, but its underlying nature is still an open question. A more detailed investigation of its properties is being pursued with more refined cosmological observations together with efforts towards direct detection with
high-precision measurements at low energies.

An important class of dark-energy theories that can naturally address the so-called cosmic coincidence problem are \emph{quintessence} models \cite{Ratra1988,Caldwell1998,Zlatev1999} involving a scalar field with particular kinds of potentials, such as inverse power laws, that imply very light masses at the present cosmological epoch. However, since they are expected to couple to regular matter with a strength comparable to the gravitational interaction, they would mediate a long-range interaction acting as a ``fifth force'' that would have been detected in tests of the equivalence principle and solar system measurements.
Interestingly, \emph{chameleon} models \cite{Khoury2004} can evade detection through a screening mechanism: whenever the matter density is not too small, its coupling to the chameleon field leads to a larger effective mass of the field and the interaction that it mediates becomes a short range one. In this way, only a thin layer on the surface of macroscopic objects would source the long-range interaction and even air's density would be enough to screen it. Similar conclusions apply also to \emph{symmetron} fields \cite{Hinterbichler2010}.

It has been argued that the sensitivity to chameleon and symmetron fields could be enhanced by employing microscopic masses because in that case the interaction with the test mass would not be screened (although the screening mechanism would still apply to the source mass) \cite{Burrage2015}. Atom interferometry is therefore a natural candidate for this kind of measurements and since the experiments are performed in a vacuum chamber, any screening due to air is also absent \cite{Burrage2015,Elder2016,Chiow2020}.
Indeed, ground-based experiments with atom interferometers have substantially improved the bounds on such models \cite{Hamilton2015,Jaffe2017,Sabulsky2019}, but they are limited by the fact that the atoms are freely falling.

\begin{figure}[ht]
	\begin{center}
		\includegraphics[width=\linewidth]{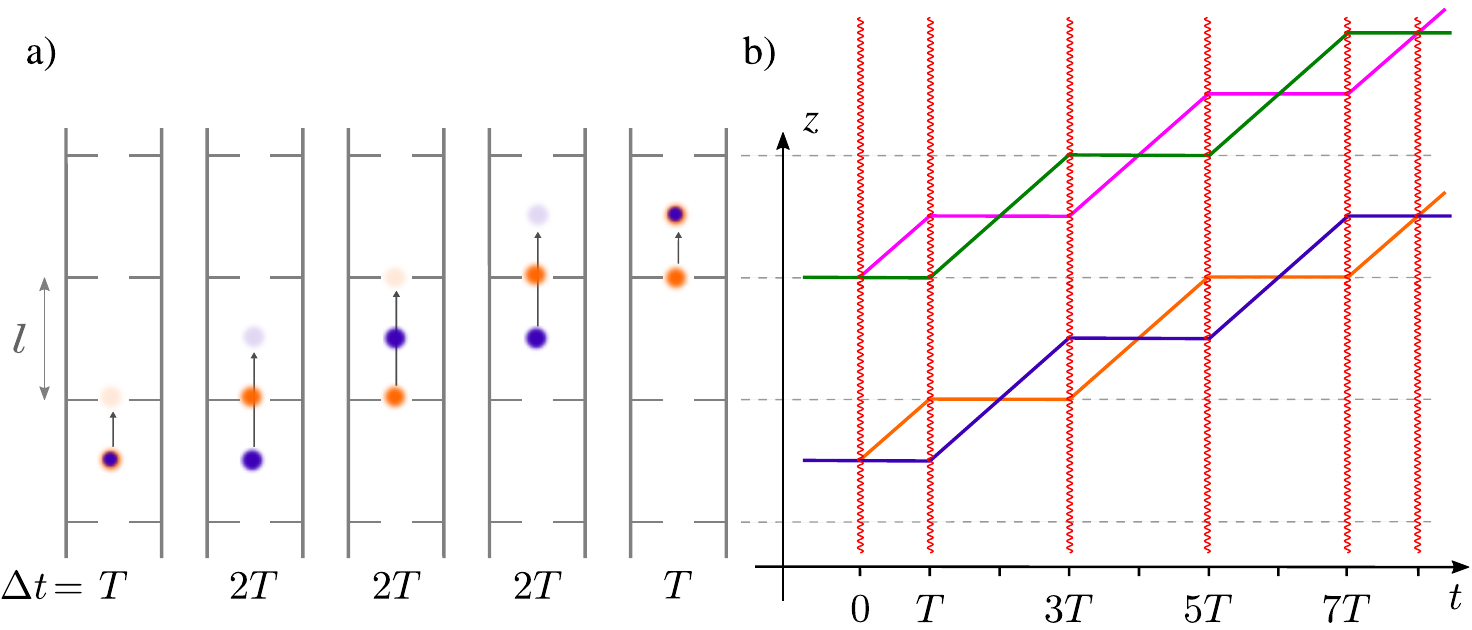}
		\caption{Scheme for dark-energy detection with atom interferometry. The left panel displays a section of the cylindrical source mass with equidistant disks separated by a distance $l$ as well as a schematic representation of the motion of the atoms in the two arms of the interferometer (blue and orange) along the symmetry axis that goes through the holes at the center of each disk.
		The right panel is a spacetime diagram of the wave-packet trajectories along the two interferometer arms. A second identical interferometer separated by a semi-integer of $l$ is simultaneously operated and interrogated by the same laser pulses. The signal of interest is obtained from a differential phase-shift measurement.
		}
		\label{fig:DarkEnergy}
	\end{center}
\end{figure}

Under microgravity conditions, on the other hand, it is possible for the atoms to spend long times close to the macroscopic source mass, which increases the atom interferometric signal associated with chameleon and symmetron fields. Nevertheless, it is crucial to mitigate the noise sources and systematic effects that also grow with time and can become very important for long interferometer times. In this respect, a particularly interesting scheme, displayed in \Fig{fig:DarkEnergy}, was proposed in Ref.~\cite{DarkEnergy2018}. The source mass is a long cylinder with a series of equidistant disks perpendicular to the cylinder walls and separated by a distance $l$. Each disk has a small hole at the center, so that the cylinder's symmetry axis, along which the atoms will move, goes through all of them as depicted in \Fig{fig:DarkEnergy}a. The scheme involves a differential measurement of two identical atom interferometers separated along the symmetry axis by a semi-integer multiple of $l$ and interrogated by common laser beams propagating along this axis. The atoms start at rest in both interferometers and a sequence of laser pulses is applied at various times as shown in \Fig{fig:DarkEnergy}b.

In addition to the initial and final beam-splitter pulses, all the intermediate mirror pulses guarantee that the maximal spatial separation between the interferometer arms, which is proportional to $T$, remains relatively small even for long total interferometer times: one simply needs to consider a large number $n$ of intermediate pulses, which leads to a total time $2n\, T$. The small arm separation and the periodic configuration of the arm trajectories and laser pulses minimize the impact of laser wave-front distortions and inhomogeneities of the magnetic field (which give rise to spurious forces through the quadratic Zeeman effect) because their effects are then similar on both interferometer arms.

The loss of contrast and systematic effects caused by rotations and gravity gradients, discussed in \Secref{Mitigation techniques}, can be naturally overcome with the mitigation techniques presented there. Indeed, by considering interferometers with an even number of loops in the spacetime diagram of \Fig{fig:DarkEnergy}b (corresponding to an even number of laser pulses), the effects of rotations with a constant rotation rate cancel out, as pointed out in \Secref{Rotations}. Similarly, the effects of gravity gradients can be compensated by adjusting the momentum transfer of all the intermediate pulses according to \Eq{eq:gg_compensation}, as explained in \Secref{Gravity gradients}.
Furthermore, the effects of laser phase noise, vibration noise or any residual accelerations of the experimental set-up, will cancel out in the differential measurement of the two interferometers shown in \Fig{fig:DarkEnergy}b, which are operated simultaneously and interrogated by common laser pulses.

To calculate the phase shift for this kind of interferometers, one can proceed as follows. First, one starts by taking into account that the phase accumulated along each interferometer arm is given by the classical action evaluated along the mean trajectory of the atomic wave packet \cite{Tannoudji,Borde2,Hogan,UGRRoura} and including the effect of the laser pulses in the potential \cite{RedshiftSchleich}. The phase shift is thus given by $S[\bm{x}^{(u)}(t)] / \hbar - S[\bm{x}^{(l)}(t)] / \hbar$, where $\bm{x}^{(u)}(t)$ and $\bm{x}^{(l)}(t)$ correspond to the mean trajectories depicted in \Fig{fig:DarkEnergy}b (orange and blue lines respectively).
Next, one separates the contribution of the potential $V_\mathrm{src}$ generated by the source mass, which includes the gravitational interaction and the chameleon or symmetron field: $S[\bm{x}^{(\alpha)}(t)] = S_0[\bm{x}^{(\alpha)}(t)] + \int dt\, V_\mathrm{src} \big( \bm{x}^{(\alpha)}(t) \big)$.
Since $V_\mathrm{src}$ is rather small, it can be treated perturbatively and the trajectories $\bm{x}^{(\alpha)}(t)$, which are classical solutions of the full action, can be expanded as $\bm{x}^{(\alpha)}(t) = \bm{x}^{(\alpha)}_{(0)}(t) + \bm{x}^{(\alpha)}_{(1)}(t) + \cdots$, where $\bm{x}^{(\alpha)}_{(0)}(t)$ is the solution of the unperturbed action $S_0$.
As shown for example in Appendix~A of Ref.~\cite{Roura2006}, when evaluating the full action to first order, it is actually sufficient to employ the unperturbed solution $\bm{x}^{(\alpha)}_{(0)}(t)$ because the contributions of the perturbations $\bm{x}^{(\alpha)}_{(1)}(t)$ cancel out.
(The same conclusions can be reached with operator-based methods \cite{PAI}.)

The unperturbed actions for the two arms are identical except for the laser phases. Hence, their contribution to the phase shift reduces to $S_0 \big[ \bm{x}^{(u)}_{(0)}(t) \big] / \hbar - S_0 \big[ \bm{x}^{(l)}_{(0)}(t) \big] / \hbar = \delta\phi_\mathrm{L}$ and the total phase shift is given by
\begin{equation}
\label{eq:DE_phase}
\delta\phi = \delta\phi_\mathrm{L}
+ \int dt\ V_\mathrm{src} \left( \bm{x}^{(u)}_{(0)} (t) \right)
- \int dt\ V_\mathrm{src} \left( \bm{x}^{(l)}_{(0)} (t) \right) .
\end{equation}
When considering the differential phase shift between the two interferometers displayed in \Fig{fig:DarkEnergy}b, the term $\delta\phi_\mathrm{L}$, associated with the laser phases, cancels out and the same would happen if we had considered small residual accelerations or vibrations. Therefore, one is left with the last two terms on the right-hand side of \Eq{eq:DE_phase} multiplied by two because these two terms give the same result for the two interferometers but with the upper and lower paths exchanged.

Considering the potential $V_\mathrm{src} (z)$ along the symmetry axis and taking the origin a the center of a disk, the differential phase shift can be well approximated by $\delta\phi_2 - \delta\phi_1 = n\, T\, \big( V_\mathrm{src} (l/2) - V_\mathrm{src} (0) \big)$, where finite-size effects of the source mass have been neglected. This result reflects the fact that the atoms in one interferometer arm spend half the time at a maximum of the potential (closest to the source), whereas in the other arm they spend half of the time at a minimum.
In this way, one can measure the potential $V_\mathrm{src} (z)$, whose $z$ dependence can be determined by repeating the measurements keeping the initial position of one of the interferometers fixed but scanning the initial position of the other. This potential includes the contributions of both the chameleon (or symmetron) and the gravitational field. In order to extract the former, one needs to subtract the gravitational potential, which can be accurately modeled for a well-characterized density distribution of the source mass. Additionally, one can employ more sophisticated source designs that suppress the gravitational contribution to the differential phase shift compared to that of the chameleon (or symmetron) field \cite{DarkEnergy2018}.

Such an experiment can be naturally performed in a space mission, but given its immunity to rotations, gravity gradients and residual accelerations, it could also be carried out in Hannover's Einstein Elevator facility \cite{EinsteinElevator}, which enables a fairly high repetition rate (up to 300 launches per day). The main limitation in this case would come from the maximum microgravity time of $4\, \mathrm{s}$ for each launch.

\subsection{Gravitational waves}
\label{sec:gravitational waves}

Gravitational waves are ripples in spacetime that propagate at the speed of light and are generated by rapidly changing mass distributions such as pairs of astrophysical objects orbiting at large velocities. The first confirmed detection by the LIGO observatory half a decade ago \cite{Abbott2016} marked the birth of gravitational-wave astronomy and opened a whole new window into the universe.
Besides LIGO \cite{LIGO} and Virgo \cite{Virgo}, additional ground-based gravitational-wave detectors are being pursued in Japan, Australia and India. Furthermore, after a successful Pathfinder mission \cite{LISA-Pathfinder} the launch of LISA \cite{LISA}, a space-borne gravitational-wave observatory, is planned for 2034.

A monochromatic plane wave of gravitational radiation gives rise to the simultaneous dilation and contraction, respectively, of distances along two perpendicular directions transverse to the direction of propagation that alternate in time as the wave propagates. In gravitational-wave detectors based on laser interferometry the simultaneous dilation and contraction along perpendicular directions give rise to opposite changes in the time of flight along the two arms of a Michelson interferometer, which can be detected in the interference signal at the exit port. However, since these waves are so weak and the change in length is proportional to the arm length, very long interferometer arms are necessary. Indeed, even for LIGO's 4-km arms, the resulting change of length is still several orders of magnitude smaller than the proton radius.

\begin{figure}[h]
	\begin{center}
		\includegraphics[width=\linewidth]{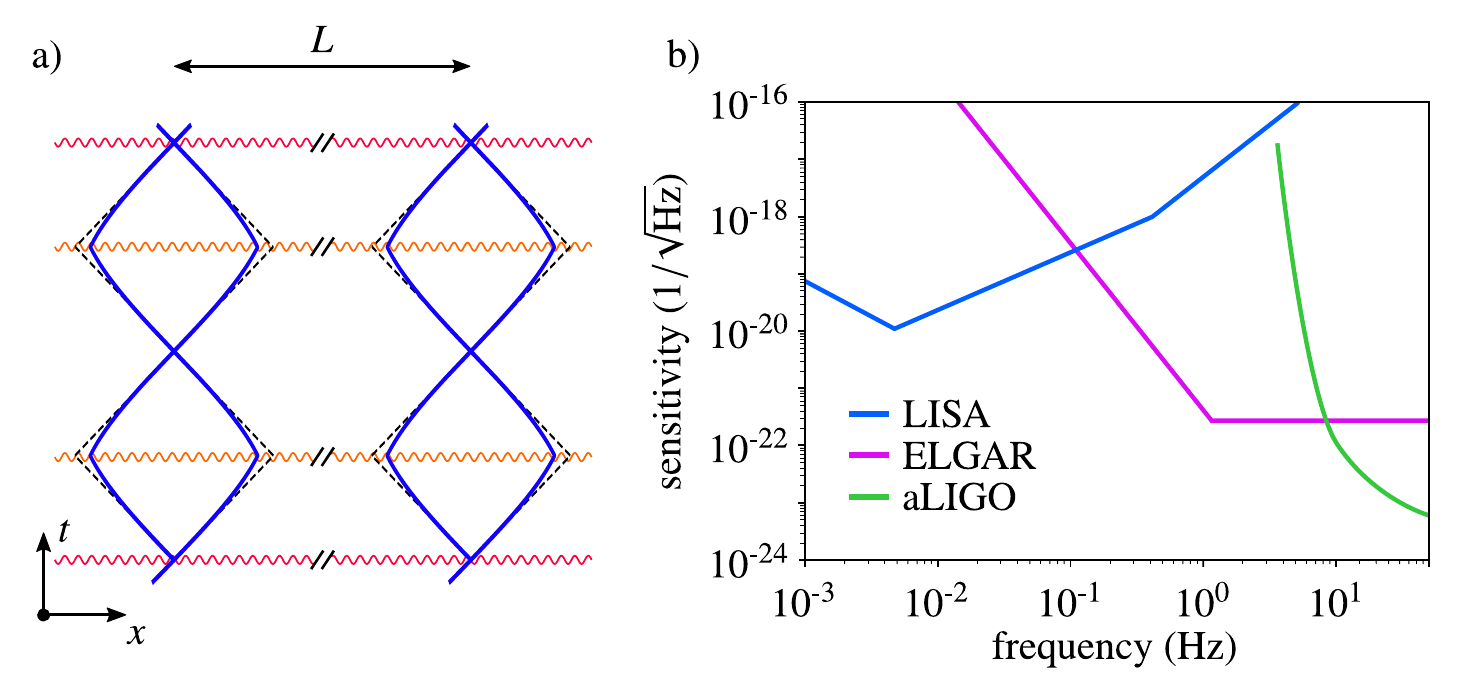}
		\caption{(a) Spacetime diagram of the wave-packet trajectories for a pair of atom interferometers separated by a long distance $L$ along the $x$ direction (not to scale) and interrogated by common laser beams propagating along this baseline. Gravitational waves affect their propagation and the phase that they imprint on the atomic wave packets, which can be read out from a differential measurement of the two interferometers. The effects of rotations and gravity gradients are respectively mitigated by using a two-loop geometry and suitably adjusting the frequency of the 2nd and 3rd pulses.
		(b) This scheme has been proposed for the European Laboratory for Gravitation and Atom-interferometric Research (ELGAR) \cite{ELGAR}, which would develop a gravitational antenna based on atom interferometry targeting the detection of gravitational waves with frequencies lying between the frequency ranges accessible to LIGO/Virgo and LISA, as shown by the strain sensitivity curves plotted in the right panel.	}
		\label{fig:GravWaveDetect}
	\end{center}
\end{figure}

Gravitational waves also affect the propagation of matter waves and it was argued in Ref.~\cite{Chiao2004} that fundamental differences compared to light propagation could be exploited to build table-top atom interferometers with sensitivities comparable to LIGO. A more careful analysis, however, revealed that this was not the case \cite{Roura2006} and that the corrected results could be reconciled with those obtained in other studies \cite{Linet1976,Stodolsky1979,Borde2004,Tino2007}.
Much more viable proposals put forward after that \cite{Dimopoulos2008b,Hogan2011} involve pairs of atom interferometers separated by a long distance $L$ and interrogated by common laser beams propagating along the baseline as shown in \Fig{fig:GravWaveDetect}a. Similarly to laser interferometers, these proposals rely upon the effect of gravitational waves on laser light propagating along the long baseline, whereas the atom interferometers at each end act as inertial references (instead of freely suspended mirrors) and are also employed to read out the small changes in the relative laser phase $\Delta\varphi_j$ of the different pulses caused by gravitational waves;
see Ref.~\cite{Review_GravWaves_geiger2017} for a review.

The potential advantages of these proposals are connected with the use of atom interferometers as inertial references. Firstly, in contrast to freely suspended mirrors, atom interferometers do not suffer from (low-frequency) suspension noise.
Secondly, besides the atom interferometers at the two ends of the baseline, a number of additional atom interferometers distributed along the baseline can be included to help discriminate and subtract the contributions of Newtonian noise \cite{Chaibi2016,ELGAR,Canuel2020b}.
Both suspension noise and Newtonian noise are responsible for the limited sensitivity at tens of Hz and below of ground-based laser-interferometric gravitational-wave detectors.
Therefore, atom-interferometric gravitational antennas are a potential candidate for detecting gravitational waves whose frequencies lie between the two frequency ranges accessible to LISA and LIGO, as shown in \Fig{fig:GravWaveDetect}. 

On the other hand, one of the main challenges for this kind of detectors is the low flux of ultracold atoms that can be currently achieved in atom interferometers compared to the much higher flux of photons in laser interferometers such as LIGO or Virgo, which drastically limits the phase-shift resolution that can be achieved by the former. Besides increasing the atom flux, quantum metrology techniques making use of entangled atoms can be exploited to push the phase-shift resolution limit several orders of magnitude below the shot-noise limit $1/\sqrt{N_\mathrm{at}}$, with the ultimate limit being set by the Heisenberg limit, which scales as $1/N_\mathrm{at}$. However, these techniques are still at a rather early development stage. Another strategy is to enhance the atom-interferometric phase shift caused by the gravitational wave by replacing each one of the diffraction pulses with a sequence of many pulses, possibly combined with the use of higher diffraction orders. The resulting phase shift scales with the total momentum transfer $k_\mathrm{eff}$, which is proportional to the number of pulses in the pulse sequence, and momentum transfers corresponding to thousands of pulses or more are being considered \cite{ELGAR}.
(Note, however, that laser interferometers on ground also get an enhancement factor from the storage time in the Fabry-Perot cavities, which roughly corresponds to bouncing back and forth several hundred times between the two mirrors.)

For the reasons mentioned above, atom-interferometric gravitational antennas are far from being competitive with the shot-noise limited sensitivity of laser interferometers. Nevertheless, their potential domain of applicability could be the detection of gravitational waves with low frequencies in the Hz and dHz regime, where the sensitivity of laser interferometers on ground is severely limited by suspension noise and Newtonian noise and is orders of magnitude worse than the shot-noise limit. 
Furthermore, in order to reduce aliasing effects due to the finite sampling rate, it is necessary to operate concurrently a number of interferometers with atom clouds launched at a rate significantly higher than the targeted detection frequencies, and this is easier for lower frequencies.

Alternatively, for space detectors with very long baselines of millions of kilometers or more, the photon flux at the receiver is much smaller and feasible atom fluxes are no longer a major limitation. Therefore, atom-interferometric gravitational antennas using heterodyne laser links \cite{Hogan2016} and targeting gravitational-wave frequencies in the mHz regime and below could eventually be competitive with laser interferometers. Moreover, by employing the atom-interferometry scheme based on single-photon clock transitions \cite{Graham2013} discussed below, such a space-borne gravitational-wave detector could be implemented with a single baseline and without the need for drag-free satellites, which could significantly reduce the complexity and costs of the mission.
Maximizing the sensitivity to such low frequencies implies rather long interferometer times of several hundred seconds, which require extremely narrow momentum widths \cite{LorianiSpeciesSelection} and the use of atomic lensing techniques, discussed in \Secref{Atomic lensing}, to achieve that.

Reaching phase-shift sensitivities in the $\mu\mathrm{rad}$ regime with a $k_\mathrm{eff}$ corresponding to thousands of photon recoils will require a great control of laser intensity fluctuations and  wave-front distortions. Other important noise sources are the coupling of initial-position and -velocity jitter to rotations, gravity gradients and wave-front distortions \cite{Canuel2020b}. Mitigation techniques addressing the effects of rotations and gravity gradients will be discussed in \Secref{Mitigation techniques}. As shown in \Fig{fig:GravWaveDetect}, one possibility is to use a two-loop interferometer geometry that mitigates the effects of rotations together with a suitable frequency adjustment of the 2nd and 3rd pulses for the compensation of static gravity gradients.
This scheme was identified as the most suitable for a recently proposed large-scale facility \cite{ELGAR,Canuel2020b} involving two horizontal baselines with two beams at different heights per baseline, and could be tested in a 150-m single-baseline prototype already commissioned \cite{MIGA} in a low-noise underground laboratory. 
Alternatively, one could use three-loop geometries with different momentum transfers per pulse [$k_\mathrm{eff}$, $(9/4) k_\mathrm{eff}$ and $(5/2) k_\mathrm{eff}$] which simultaneously suppress the effects of rotations and gravity gradients \cite{Hogan2011}. Such interferometer geometries can be implemented in space and in ground experiments with a single vertical baseline. However, they cannot be employed in horizontal configurations with two beams per baseline unless the atoms are relaunched during the interferometer sequence \cite{Schubert2019}.

The proposals for gravitational-wave detection with atom interferometry mentioned above involve diffraction pulses based on two-photon transitions such as Raman or Bragg diffraction.
An alternative atom interferometry scheme using a diffraction mechanism based on single-photon transitions between the two clock states in atoms such as Sr or Yb, which are commonly employed in optical atomic clocks, was proposed in Ref.~\cite{Graham2013} (see also Ref.~\cite{Yu2011} for a related suggestion).
The great advantage of this scheme is that differential measurements with pairs of atom interferometers are immune to laser phase noise even for very long baselines and a single baseline is therefore sufficient.
This is advantageous for space missions, where it can lead to substantial simplifications \cite{Graham2013,Hogan2016}, and it enables vertical configurations for ground experiments.
In fact, projects for ground-based prototypes with a vertical 100-m baseline \cite{MAGIS,AION}, which can also be exploited to search for ultralight dark matter \cite{Arvanitaki2018}, are already underway and plans for future space missions are being considered too \cite{Graham2017,SAGE,AEDGE}.

It should be noted that the clock transition is forbidden for bosonic isotopes unless a strong magnetic field is applied, which does not seem a viable option for high-precision measurements. Hence, one will need to employ fermionic isotopes instead. For long interferometer times it is crucial to use ultracold atoms, as discussed in the next section. However, Bose-Einstein condensation is not possible for fermions and one will have to use fermionic clouds close to quantum degeneracy, which will not be discussed in this review.

\section{BECs and atomic lensing} 
\label{BEC}
The proposals of the previous sections suggest interferometer times of tenth of seconds in microgravity. While large interrogations times are associated with high sensitivity, problems related to the size of the atom cloud are implied as well.
Initially the atoms are confined by e.g.~a dipole trap. However, due to the cloud's finite momentum spread, the it will start to expand drastically after release. 
Indeed, an order of magnitude estimation for cold rubidium atoms indicates a size of several tenth of centimeters after ten seconds of interferometer time \cite{ThermalCloudBEC}. Atom clouds of large extent in position and momentum space make long-time interferometry experimentally difficult if not impossible for the following reasons:
A cloud with broad momentum distribution cannot be addressed resonantly with the laser beams, leading to velocity selectivity \cite{Bragg2} and consequently inefficient diffraction. In addition, at the exit port of the interferometer the diffraction orders cannot be discriminated if the momentum width of the cloud is of the order of $\hbar k_\mathrm{eff}$. 
In position space the Rabi frequency should be uniform over the size of the cloud for efficient diffraction, thus its spatial extent should be small compared to the beam width (of the order of a cm).
Furthermore, the small atom density of a large sample cannot be resolved accurately at the detector. Finally, large clouds are strongly susceptible to wave-front distortions \cite{WaveFront_2011,LaserPhaseNoise}.
During the diffraction process the spatially dependent phase of the effective laser beam is imprinted on the wave packet. Since the phase of a realistic laser beam deviates from the ideal planar wave fronts, phase shifts dependent on the shape of the atomic wave packet are introduced at each laser pulse. 
While this bias can be partially corrected for \cite{WaveFront_2011, WaveFront_2015},
wave-front distortions are one of the limiting systematic effects in absolute gravimetry  \cite{SYRTE,WaveFrontLimiting1} but can be significantly decreased with smaller atom clouds \cite{Wavefront_2018,ThermalCloudBEC}.

An obvious strategy to achieve smaller cloud sizes is cooling to ultra-low temperatures since
the expansion rate scales with the spread in momentum space which in turn decreases with the temperature of the sample.  In case of bosonic atoms, the system will eventually reach quantum degeneracy and undergo Bose-Einstein condensation with a momentum width given approximately by the Heisenberg uncertainty relation \cite{TFMomentumWidth}. To provide some intuition, the momentum spread of a BEC in a trap of $2\pi\cdot 50 \text{Hz}$ is comparable to the one of a Maxwell-Boltzmann distribution at the  pico-Kelvin range. 
A second strategy to achieve small cloud sizes is atomic lensing, where the clouds are collimated after some free expansion time. This method, which is described in more detail in \Secref{Atomic lensing}, is applicable to both ultra-cold non-degenerate samples and BECs. 
A combination of BECs as atom source and atomic lensing for further reducing the expansion rate,  overcomes the technical problems described above and significantly reduces the impact of wave-front distortions. 
For an overview over error sources and a comparison between the properties of ultra-cold non-degenerate samples and BECs see Ref.~\cite{ThermalCloudBEC}.

Some further intuition into the properties of BECs is provided in the subsequent section.

\subsection{Bose-Einstein condensation}
Bose-Einstein condensation was first observed experimentally in 1995 \cite{CornellWiemann_Nobel,Ketterle_Nobel}  but today is reproduced routinely in many laboratories all over the world, for a review see \cite{DalvovoReview}. What makes BECs the preferable source for interferometry in many applications compared to cold non-degenerate samples is the extraordinary coherence length and small momentum spread. 

In a thermal ensemble many modes are occupied according to the Bose-Einstein distribution. Consequently, the coherence length of this mixture is much smaller than the spatial extent of the cloud. At the critical temperature the system undergoes Bose-Einstein condensation, arising from combinatorial properties of the indistinguishable bosonic particles.
Below the critical temperature a major fraction of the atoms occupies a single mode and the coherence length becomes comparable to the width of the condensate. As a consequence of the single-mode occupation, a BEC in a harmonic trap approximately exhibits a momentum width given by the Heisenberg uncertainty relation \cite{TFMomentumWidth}.
Mathematically, the intuition given above can be formalized by the Penrose-Onsager criterion \cite{PenroseOnsager}, serving as a rigorous definition of BEC.

The ground state and time evolution of the BEC mode function is described by the Gross-Pitaevskii equation \cite{GPE1, GPE2}, which takes the form of a nonlinear Schrödinger equation. The additional self-interaction term accounts for interactions among the particles which is much more pronounced in BECs than in much more dilute thermal samples with the same number of particles. The Gross-Pitaevskii equation can be derived invoking symmetry breaking  \cite{SymmetryBreaking_Bogoliubov, SymmetryBreaking_Yukalov} or number-conserving approaches  \cite{NumberConserving1, NumberConserving2}. 

Light-pulse interferometers with BECs as atom source \cite{Bragg2,BEC_Gravimetry2011,Bragg4}  have shown nearly perfect contrast \cite{BEC_Gravimetry2011}. BECs have been employed in microgravity \cite{VanZoest2010,Muntiga2013} reaching several seconds of free-evolution time and are a basic element of proposals for long-time interferometry testing fundamental properties of gravity.

\subsection{Atomic lensing}
\label{Atomic lensing}
\begin{figure}[H]
	\begin{center}
		\includegraphics[width=\textwidth]{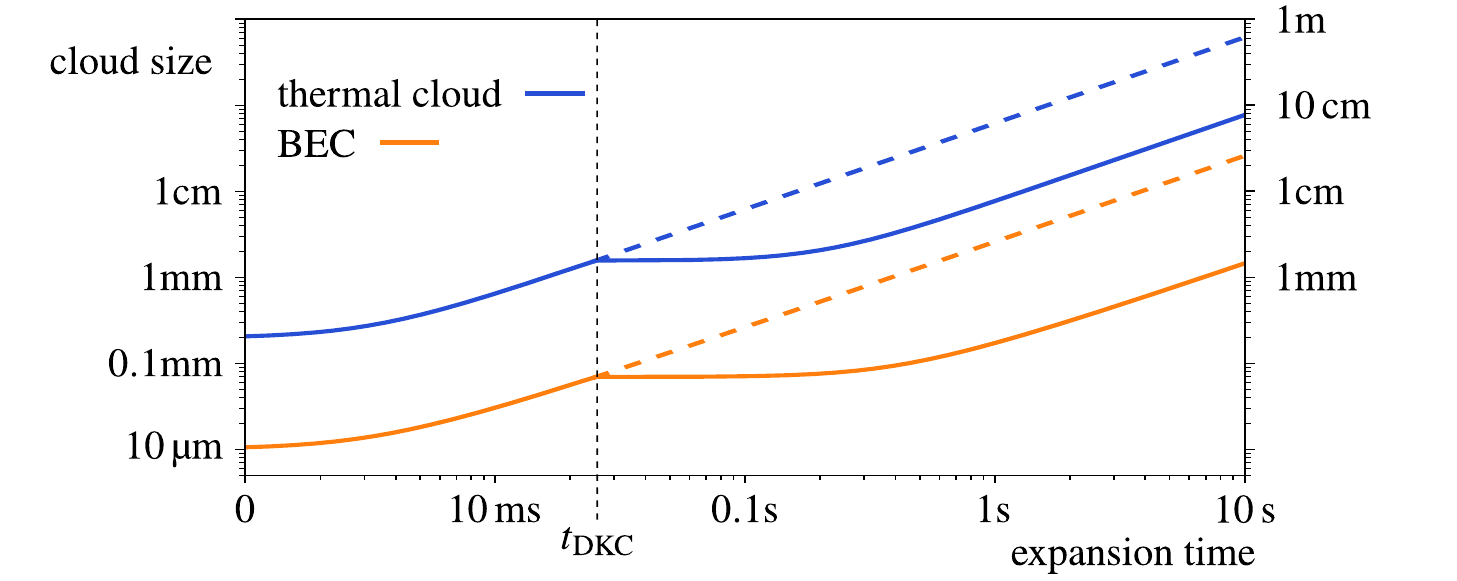}
		\caption{Comparison of the expansion dynamics of a BEC and a thermal cloud above the critical temperature. Initially a zero-temperature BEC and a thermal cloud with a temperature of $50\text{\textmu K}$ are confined in a harmonic trap with frequency $\omega=2\pi\cdot 50\text{Hz}$. Upon release both states start to expand. After the initial free evolution time  of $t_\mathrm{DKC}=25$ms the expansion is decelerated by delta-kick collimation.
		The large size of about $10$cm of the thermal state after $10$s expansion indicates a clear advantage of using BECs for long-time interferometry which reaches a final cloud size of only $1$mm.
		The parameters for each state are chosen for $10^6$ $^{87}\mathrm{Rb}$ atoms.}
		
		\label{fig:AtomicLensing}
	\end{center}
\end{figure}

\Fig{fig:AtomicLensing} compares the expansion rates after release from a harmonic trap between a BEC and a thermal cloud. As can be seen from the figure, the expansion rate of the BEC is much smaller than the one of the thermal ensemble but it still reaches a size of several centimeters after an expansion time of 10 seconds (dashed lines in the figure). To reach experimentally acceptable final cloud sizes, 
a collimation step known as delta-kick collimation \cite{DKC_1997, DKC_1999} or alternatives \cite{MatterWaveLensing2015} is applied. After an initial free expansion time, the atom trap is switched on again for a specific amount of time. By that, kinetic energy is extracted from the system and the subsequent expansion can be drastically reduced.
Theoretically, the correct timing and duration of the delta-kick pulse can be calculated using approximate scaling solutions available for the Gross-Pitaevskii equation \cite{ScalingApproach1,ScalingApproach2,ScalingApproach3} and the Liouville equation \cite{ScalingLiouville} that can be used to model a thermal non-degenerate quantum gas. Today, delta-kick collimation is routinely applied in experiments on ground \cite{DeltaKick2013, AtomChipGravi} and in drop-tower facilities \cite{Muntiga2013} in microgravity.

Experiments involving two different atomic species, as for example in tests of the WEP, are sourced by mixtures of BECs which are first cooled to a common ground-state configuration 
\cite{Bose_BoseMixture, ClassificationBosBose1,ClassificationBosBose2}.
Alternatively, a mixtures of a BEC and a degenerate Fermi gases \cite{Bose_Fermi_Mixture1} could be employed for which a rich class of groundstate configurations has been predicted \cite{Classification_Bose_Fermi}.
In case of different trapping frequencies of the species and the resulting different expansion rates after release, more sophisticated collimation techniques are required.
For Boson-Boson mixtures particularly tailored collimation sequences \cite{TwoSpeciesDKC} can achieve small expansion rates for both components.

\subsection{Interaction-related systematics}
The collimated small atom clouds with relatively high atom density bring about new parasitic effects arising from particle-particle interaction.
As the interaction strength quickly decreases with the size of the cloud,  an initial expansion time before the interferometer sequence and the collimation step \cite{STEQUEST} is crucial.
There are two negative impacts of particle-particle interactions briefly explained in the following. The first is referred to as mean-field phase shifts which are captured by the non-linear term in the Gross-Pitaevskii equation. In case of a particle imbalance between the branches a differential systematic phase shift is accumulated \cite{Clade2015}. Therefore, in ambitious tests of gravity, high control over the beam-splitting process is necessary \cite{STEQUEST}.
Mean-field phase shifts are also introduced during separation of the cloud that follows the beam splitter \cite{MeanFieldVelocity}. On a theoretical level, BECs in interferometers with large branch separation can be treated from the perspective of observers comoving \cite{Gupta2014, UfrechtPhd} with the clouds on the interferometer branches. By that, the system becomes numerically tractable.
Note that interaction effects could also be canceled by tuning the particle-particle interaction strength  \cite{TuningScattering} via Feshbach resonances. However, the fluctuations of the required strong magnetic fields would lead to large systematic effects. 

The second detrimental effect is due to the quantum state after the first beam splitter which is a superposition of different particle-number states distributed over the two branches of the interferometer with fixed total number of atoms. The fluctuation of size $\sqrt{N}$ of the distribution after the first beam splitter is responsible for the shot noise in case of non-interacting particles. In case of interacting BECs the mean-field phase shift depends on the atom density, thus each of the states in the superposition acquires a different phase shift. The resulting effect is a derogation of the relative phase between the branches and is known as phase diffusion \cite{leggett1991, LewebsteinYou, JavanainenWilkens, Wright1996,sinatra2000} which leads to larger atom-number fluctuations after the final beam splitter. An estimation of this effect in the Thomas-Fermi limit \cite{sinatra2000} shows that its impact can be disregarded for small interferometer times \cite{BEC_Gravimetry2011} but drastically reduces the single-shot sensitivity in long-time interferometry \cite{UfrechtPhd}.

\section{Mitigation techniques}
\label{Mitigation techniques}

A relative displacement between the mean positions of the two interfering wave packets at each exit port, or a difference between their mean momenta, leads to a reduction of the interferometer contrast in \Eq{eq:AI_probability}, which lowers the strength of the interference signal and the sensitivity of the interferometer \cite{OvercommingLoss}. Furthermore, such displacements between the interfering wave packets also imply a dependence of the interferometer phase shift on the mean position and velocity of the initial state \cite{OvercommingLoss,Roura2017}.
Both rotations and gravity gradients give rise to open interferometers with non-vanishing relative displacements that grow with the interferometer time and pose a major challenge for long-time atom interferometry. Indeed, the resulting sensitivity to the initial kinematics of the atomic wave packet is a major source of systematic effects that imposes very demanding requirements on the control of the initial position and velocity in high-precision measurements.
Mitigation techniques that successfully address the effects associated with rotations and gravity gradients will be presented in the next two subsections. In addition, we will discuss in a third subsection how orbital modulation can be exploited to mitigate these and other systematic effects in space-based atom-interferometry measurements.

    \subsection{Rotations}
    \label{Rotations}

Describing the evolution of the atomic wave packets is simpler in a non-rotating frame, but then one needs to take into account the rotations of the experimental set-up and the laser beams with respect to that frame \cite{Kleinert}; in particular, for ground experiments the laboratory frame will be co-rotating with the Earth. As a consequence, the momentum transfer $\bm{k}_\mathrm{eff}$ associated with each laser pulse will be rotated by a different angle and to lowest order amounts to a change $(\bm{\Omega}\, t_n) \times \bm{k}_\mathrm{eff}$ for the $n$th laser pulse, where $\bm{\Omega}$ is the angular velocity and $t_n$ is the time at which the pulse is applied. Altogether the changes for the pulses in a MZ interferometer lead to relative displacement between the mean positions of the interfering wave packets perpendicular to $\bm{k}_\mathrm{eff}$ and given by $-2 (\bm{\Omega}\, T) \times (\bm{k}_\mathrm{eff} / m)\, T$ \cite{Kleinert,OvercommingLoss}. This displacement implies a dependence of the phase shift on the initial velocity according to \Eq{eq:rotation_phase}. Moreover, for sufficiently long interferometer times (or large angular velocities) the mismatch between the interfering wave packets can cause a substantial loss of contrast. In what follows, we present two mitigation techniques that successfully overcome these difficulties.

(i) One method for compensating the effects of rotations relies on the use of a \emph{tip-tilt mirror} for retro-reflection of the laser beams and works as follows. The two counter-propagating laser beams with wave vectors $\bm{k}_1$ and $\bm{k}_2$ in \Fig{fig:RamanBraggMZ}, which give rise to a total momentum transfer $\bm{k}_\mathrm{eff} = \bm{k}_1 - \bm{k}_2$, are obtained by injecting two co-propagating beams with frequencies $\omega_1$ and $\omega_2$ that are reflected back by a retro-reflection mirror. Bragg or Raman diffraction can be understood as absorption of a photon with momentum $\bm{k}_1$ from the incoming beam and the stimulated emission of another photon with momentum $\bm{k}_2$ induced by the reflected beam with frequency $\omega_2$. The two incoming beams are parallel with wave vectors $\bm{k}_1 \parallel \bm{k}'_2$, and for a mirror perpendicular to them one has $\bm{k}_2 = - \bm{k}'_2$ for the reflected beam. However, if one slightly tilts the mirror so that it is no longer exactly perpendicular to the incoming beams, one has $\bm{k}_2 \neq - \bm{k}'_2$, but the resulting $\bm{k}_\mathrm{eff}$ is still perpendicular to the mirror. Hence, by tilting the mirror by a small angle, one can change the direction of $\bm{k}_\mathrm{eff}$. In particular, one can choose the tilt angle (and direction) for every pulse so that it compensates the effect of rotations. Indeed, although rotations of the experimental set-up would lead to a change of direction of the incoming beams with respect to the non-rotating frame, tilting the mirror accordingly in each case so that its orientation remains the same for all pulses guarantees that the direction of $\bm{k}_\mathrm{eff}$ is left unchanged in this frame for the whole interferometer sequence.

The use of a tip-tilt mirror for compensation of rotations, which was proposed in Ref.~\cite{Hogan} and experimentally implemented in Refs.~\cite{Lan2012,FirstPublicationStanfordTower}, has by now become a standard technique.

(ii) A second method for compensating the effects of rotations involves employing atom interferometers with \emph{multi-loop geometries} \cite{Marzlin,Dubetsky2006,Kleinert}. The simplest case is the two-loop interferometer consisting of 4 pulses shown in \Fig{fig:GravWaveDetect}. For the choice of times between pulses displayed there and corresponding to $(T, 2T, T)$ the contributions to the relative displacement between interfering wave packets cancel out to first order in $(\bm{\Omega}\, T)$ \cite{Marzlin,Dubetsky2006,Kleinert}.
This interferometer geometry has been employed in applications to high-precision measurements of the rotation rate \cite{Stockton2011,Gyroscopy2016,Gyroscopy2018} that benefit from the insensitivity to the initial velocity $\bm{v}_0$ and the absence of contrast loss due to rotations despite the long interferometer times.
Indeed, the best cold-atom gyroscopes to date, with a long-term stability of $3 \times 10^{-10}\, \mathrm{rad/s}$, rely on this kind of interferometers \cite{Gyroscopy2018}.
Note that the phase-shift contributions of the same form as the right-hand side of \Eq{eq:rotation_phase} cancel out in this case and cannot be used to determine the rotation rate. Nevertheless, there is still the following contribution that depends on the angular velocity:
\begin{equation}
\label{eq:rotation_phase2}
\delta\phi = 4\, \bm{k}_\mathrm{eff} \cdot \left(\bm{\Omega} \times \bm{g}  \right)T^3 .
\end{equation}
Therefore, when combined with an independent measurement of the local gravitational acceleration $\bm{g}$, the projection of the angular velocity on the direction perpendicular to $\bm{k}_\mathrm{eff}$ and $\bm{g}$ can be inferred.

In atom interferometers with multi-loop geometries that compensate the effects of rotations the contributions to the leading term associated with the gravitational acceleration, proportional to $\bm{k}_\mathrm{eff} \cdot \bm{g} \, T^2$, cancel out too \cite{Marzlin}. Thus, this kind of interferometers are not suitable for gravimetric measurements or UFF tests. On the other hand, they can still be useful for gradiometric measurements of small time-dependent variations of the gravitational field with frequencies comparable to $1/T$, which have interesting geophysical applications and are also being considered as gravitational antennas for the detection of gravitational waves in the infrasound regime \cite{ELGAR,Canuel2020b}, as shown in \Fig{fig:GravWaveDetect}.
It can also be applied to the dark-energy measurements discussed in \Secref{sec:dark_energy} and illustrated in \Fig{fig:DarkEnergy} as long as an even number of loops is employed.

    \subsection{Gravity gradients}
    \label{Gravity gradients}
    
Deviations from a uniform gravitational field can be characterized in terms of the gravity-gradient tensor $\Gamma_{ij} = - \partial^2 U / \partial x^i  \partial x^j$, where $U$ is the gravitational potential, and give rise to tidal forces $f_i = \Gamma_{ij} \,x^j$. Assuming spherical symmetry, the gravity-gradient tensor associated with Earth's gravitational field is given by $\Gamma = \mathrm{diag} (\Gamma_{xx}, \Gamma_{yy}, \Gamma_{zz}) = (GM_\oplus / 2 r^2) \, \mathrm{diag} (-1,-1,2)$, where $z$ corresponds to the radial direction and $x, y$ lie on the orthogonal plane. Near Earth's surface $\Gamma_{zz} \approx 3 \times 10^{-6}\, \mathrm{s}^{-2}$.

\begin{figure}[ht]
	\begin{center}
		\includegraphics[width=\linewidth]{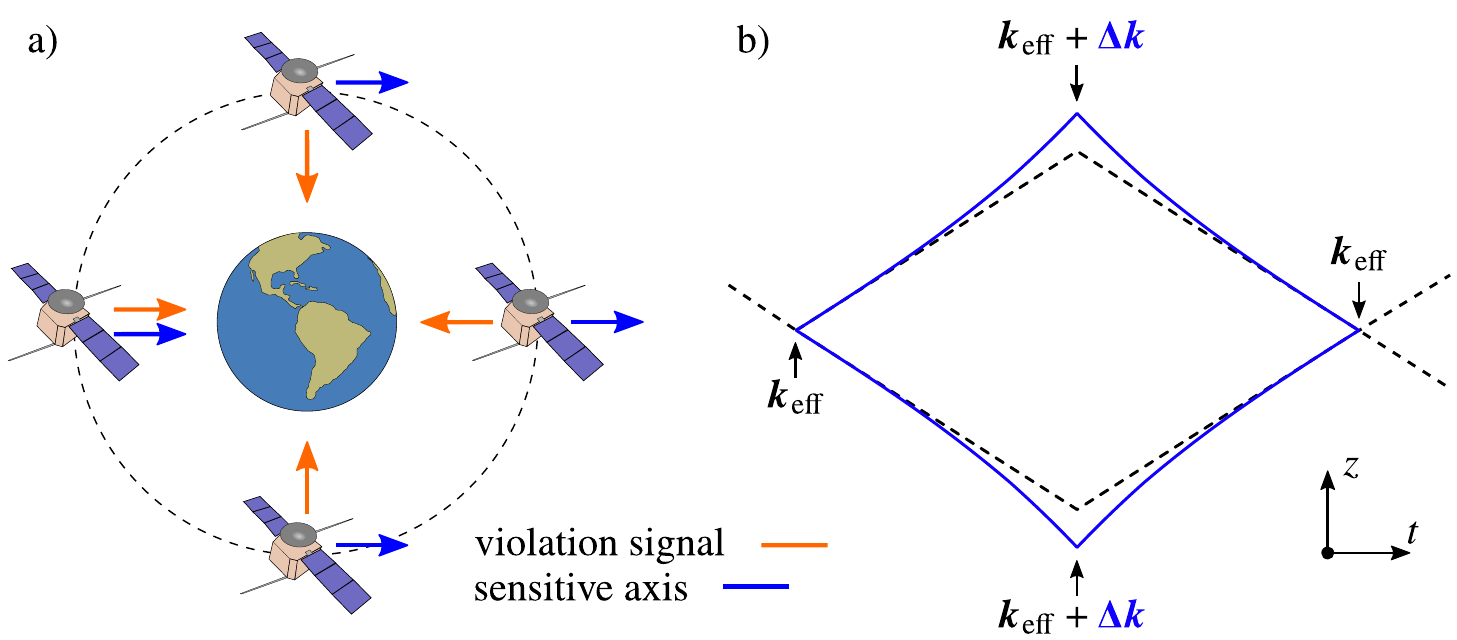}
		\caption{Orbital modulation (a) and  gravity-gradient compensation (b) are two examples of effective mitigation techniques.
		(a) In an inertial-pointing spacecraft $\bm{k}_\mathrm{eff}$, and hence the sensitivity axis, remains constant with respect to a non-rotating frame, whereas the direction of the local gravitational acceleration $\bm{g}$ changes along the orbit. The signal of interest (possible UFF violations) is therefore modulated with a frequency $\omega_\mathrm{orb}$ and can be separated from systematic effects whose modulation frequencies are different multiples of $\omega_\mathrm{orb}$.
		(b)~Spacetime diagram of the wave-packet trajectories in an atom interferometer for a freely falling frame. For a uniform field these trajectories would be straight lines (dashed), but tidal forces open them up (continuous lines) and would lead to an open interferometer. This can be overcome by suitably adjusting the momentum transfer from the central pulse.}
	\label{fig:sfs}
	\end{center}
\end{figure}

In order to understand the effects of gravity gradients in an atom interferometer, it is convenient to consider a suitable freely falling frame, as shown in the spacetime diagram of \Fig{fig:sfs}b. In this frame the atomic wave packets would follow straight lines (dashed lines) if the field were uniform, but the tidal forces tend to open up the trajectories (continuous lines) and lead to and open interferometer. The position and momentum mismatch between the interfering wave packets can cause a substantial loss of contrast for long interferometer times and gives rise to the following phase-shift contributions that depend on the initial position and momentum \cite{Roura2017,OvercommingLoss}:
\begin{equation}
\label{eq:gg_phase}
\delta\phi = \bm{k}_\mathrm{eff}^\mathrm{T} \left( \Gamma\,T^2\right) \big( \bm{x_0} + \bm{v}_0 T \big) ,
\end{equation}
where the superindex $\mathrm{T}$ denotes matrix transposition.
As a result, small discrepancies in the initial kinematics of the two atomic species can mimic violations of the equivalence principle in tests of UFF.
For example, reaching a precision of $\eta_{1,2} \lesssim 10^{-15}$ would require a control of the relative initial position of the two atomic species at the level of $1\, \mathrm{nm}$ and of the initial relative velocity at the level of $0.1\, \mathrm{nm/s}$ for $T = 5\, \mathrm{s}$, which are extremely demanding requirements.
The sensitivity to the initial kinematics of the two atomic clouds is also the main systematic effect in measurements of the gravitational constant $G$ and an important aspect driving the design of the experimental set-up in order to minimize it \cite{Newton3,Rosi2018}.

A very effective technique for overcoming these difficulties was proposed in Ref.~\cite{Roura2017}. The key idea is to change slightly the momentum transfer of the second pulse by
\begin{equation}
\label{eq:gg_compensation}
\Delta \bm{k} \approx \frac{1}{2} \left( \Gamma\, T^2 \right) \bm{k}_\mathrm{eff} ,
\end{equation}
so that for each arm the trajectory of the atomic wave packet after the central pulse is the time reversal of the trajectory before the pulse, as illustrated in \Fig{fig:sfs}b, and one is left with a closed interferometer with no relative displacement between the interfering wave packets.
This change of the momentum transfer can be accomplished by increasing the two frequencies $\omega_1$ and $\omega_2$ appearing in \Fig{fig:RamanBraggMZ} by $\Delta k / c$, which amounts to increasing the detuning $\Delta$ with respect to the auxiliary state while keeping the resonance condition for the two-photon process.

The experimental implementation of this method for gravity-gradient compensation was demonstrated in Refs.~\cite{d_Amico2017,Overstreet2018} and has been instrumental in improving by four orders of magnitude the bounds on violations of UFF based on atom interferometry \cite{WEPKasevich}, with prospects for further improvement in the near future. Moreover, by exploiting this technique, it should be possible to perform new measurements of $G$ that are competitive with the best measurements using macroscopic test masses and even outperform them \cite{d_Amico2017,Rosi2018}.
It is also an important ingredient in the search of certain kinds of dark-energy using atom interferometry proposed in Ref.~\cite{DarkEnergy2018} and illustrated in \Fig{fig:DarkEnergy}.
Finally, the technique can also be employed to compensate the effects of static gravity gradients in gravitational antennas as illustrated in \Fig{fig:GravWaveDetect}, where the effect of the tidal forces is the opposite of \Fig{fig:sfs}b because Earth's gravity gradient along the vertical and horizontal directions have opposite sign.

Note that while the method outlined above applies to quadratic gravitational potentials (i.e.\ approximately uniform gravity gradients), it can be straightforwardly extended to anharmonic potentials by changing appropriately the frequencies
of two laser pulses instead of one \cite{Overstreet2018,Ufrecht2021}.

     \subsection{Orbital modulation}
     \label{Orbital modulation}

In laboratories on Earth the experimental set-up typically co-rotates with Earth. In contrast, experiments in orbit around Earth offer more flexibility in this respect. In fact, one can take advantage of the orbital modulation to separate the signal of interest from various kinds of systematic effects \cite{Chiow2017,Loriani2020}. A simple choice is a spacecraft with inertial pointing, i.e.\ vanishing angular velocity with respect to the non-rotating frame, where the direction of $\bm{k}_\mathrm{eff}$ is fixed and unwanted effects associated with rotations are minimized. For a UFF test with a differential measurement of two atomic species, as described in \Secref{Tests of the weak equivalence principle}, the signal is given by $\bm{k}_\mathrm{eff} \cdot (\bm{g}_2 - \bm{g}_1 )\, T^2 = \eta_{2,1}\, \bm{k}_\mathrm{eff} \cdot \bm{g}\, T^2$ and  for a circular orbit with inertial pointing $\bm{k}_\mathrm{eff} \cdot \bm{g} \propto \cos (\omega_\mathrm{orb} t)$. Instead, the systematic effects due to Earth's gravity gradient and given by \Eq{eq:gg_phase} are proportional to $\cos (2\, \omega_\mathrm{orb} t)$, as can be seen from the fact that the gravity gradient tensor is the same for any pair of antipodal points on the orbit. On the other hand, the contribution from the gravity gradient caused by the spacecraft itself is constant along the orbit and the same is true for higher orders in a multipole expansion of the gravitational field around the location of the atoms%
\footnote{Earth's gravitational field gives a much smaller contribution to higher orders in this multipole expansion which does change along the orbit, but is typically suppressed by seven orders of magnitude or more compared to the gravity-gradient term.}, which are dominated by the local mass distribution (as explained for instance in Appendix~A of Ref.~\cite{UGRRoura}).

It is therefore possible to separate those contributions with different modulation frequencies when demodulating the measured signal through Fourier analysis. In particular, while the signal of interest is modulated at a frequency $\omega_\mathrm{orb}$, systematic effects due to Earth's gravity gradient are modulated at $2\, \omega_\mathrm{orb}$ and those associated with the spacecraft's local mass distribution are time-independent (zero frequency).
In addition, an elliptic (rather than circular) orbit as well as multipoles of Earth's gravitational field characterizing the deviations from spherical symmetry will lead to small contributions with modulation frequencies corresponding to higher multiples of $\omega_\mathrm{orb}$.
This separation of the signal from systematic effects with different orbital modulation frequencies has been successfully employed by the MICROSCOPE mission \cite{MICROSCOPE_2017,Microscope_2019}, which tested the UFF at the $10^{-14}$ level with a pair of freely falling macrosopic masses in orbit.

Besides systematic effects associated with gravity gradients, one can exploit the orbital modulation to address other systematic effects too, such as those that are entirely connected with the experimental apparatus and can be maintained stable for times longer than the orbital period. In particular, this could include systematic effects associated with small gradients in the black body radiation \cite{BBR2,BBR1} provided that it is not affected by changes of the solar radiation along the orbit.

Interestingly, these demodulation methods can be combined with the gravity-gradient compensation technique proposed in Ref.~\cite{Roura2017} and described in the previous subsection. In fact, it has been argued that the systematic effects associated with gravity gradients could then be suppressed to the $10^{-18}$ level \cite{Loriani2020}.

\section{Outlook}
\label{Outlook}

Atom interferometers have enabled the most accurate measurements of the fine structure constant to date \cite{FineStructure3,Morel2020}, but have also shown their great potential in fundamental tests of gravity. Indeed, after improving the results of previous experiments by several orders of magnitude, tests of UFF based on atom interferometry have recently reached relative uncertainties of the order of $10^{-12}$ \cite{WEPKasevich} and with prospects of further improvement in the near future, which brings them very close to the best results obtained on ground experiments with macroscopic masses. Furthermore, plans for future space missions exploiting the mitigation techniques described in \Secref{Mitigation techniques} and striving for sensitivities at the $10^{-17}$ level are currently under consideration \cite{WEPSatelliteProposal}.
Similarly, atom interferometers have been employed in measurements of the gravitational constant $G$ with an accuracy of 150~p.p.m.\ \cite{Newton3}, and it has been convincingly argued \cite{Rosi2018} that new designs relying on the gravity-gradients compensation technique \cite{Roura2017,d_Amico2017} of \Secref{Gravity gradients} will become competitive with the best measurements using macroscopic test masses and even outperform them.

In addition, atom interferometers constitute an ideal tool for constraining dark-energy models that exhibit screening mechanisms capable of evading detection with macroscopic test masses. In fact, they have already contributed to excluding large parts of their parameter space \cite{Hamilton2015,Jaffe2017,Sabulsky2019}, but considerable improvements are expected by performing the experiments in microgravity \cite{DarkEnergy2018}, where the atoms can spend much longer times close to the source mass.

Finally, promising proposals for detection of low-frequency gravitational waves and searches for ultralight dark matter have not reached the same level of maturity yet, but ground prototypes with 100-m baselines \cite{MAGIS,AION} are already being commissioned. Moreover, longer-term plans for dedicated space missions \cite{Graham2017,SAGE,AEDGE} are under consideration as well.

\section*{Acknowledgements}
We thank E.~Giese, F.~Di Pumpo and A.~Friedrich for interesting discussions
and the whole QUANTUS consortium for a fruitful collaboration on these topics over the years.

\section*{Disclosure statement}
No potential conflict of interest was reported by the authors.

\section*{Funding}
This work is partly supported by the German Aerospace Center (Deutsches Zentrum f\"ur Luft- und Raumfahrt, DLR) with funds provided by the Federal Ministry of Economic Affairs and Energy (Bundesministerium f\"ur Wirtschaft und Energie, BMWi) due to an enactment of the German Bundestag under Grants No. 50WM1556 (QUANTUS IV) and 50WM1956 (QUANTUS V).
The work of IQ\textsuperscript{ST} is financially supported by the Ministry of Science, Research and Art Baden-W\"urttemberg (Ministerium f\"ur Wissenschaft, Forschung und Kunst Baden-W\"urttemberg).
W.~P.~S. is grateful to Texas A\&M University for a Faculty Fellowship at the Hagler Institute for Advanced Study at Texas A\&M University and to Texas A\&M AgriLife for the support of this work.

\end{document}